\documentclass[11pt,a4paper,english]{article}
\usepackage{titling}
\usepackage[affil-it]{authblk}
\usepackage{longtable}
\usepackage{lscape}
\usepackage{graphicx}
\usepackage{epsfig}
\usepackage{dsfont}
\usepackage{amsfonts}
\usepackage{amsmath}
\usepackage{amsthm}
\usepackage{amssymb}
\usepackage{bbold}
\usepackage[english]{babel}
\usepackage[utf8]{inputenc}
\usepackage[left=0.8in,right=0.8in,top=1.0in,bottom=1.2in]{geometry}
\usepackage{calc}
\usepackage{cancel}
\usepackage{float}
\usepackage{slashed}
\usepackage{mathrsfs}
\usepackage{pdfpages}
\usepackage{tabu}
\usepackage{simplewick}
\usepackage[hidelinks]{hyperref}
\usepackage{bm}
\usepackage{multirow}
\usepackage{hhline}
\usepackage{pifont}
\usepackage{float}
\usepackage[caption = false]{subfig}
\usepackage{xcolor}
\usepackage[normalem]{ulem}
\usepackage{cite}
\usepackage{braket}

\usepackage{datetime}
\usepackage{booktabs}

\date{}

\numberwithin{equation}{section}
\numberwithin{figure}{section}
\numberwithin{table}{section}

\makeatletter
\g@addto@macro\bfseries{\boldmath}
\makeatother

\allowdisplaybreaks

\newcommand{\mLb}{m_{\Lambda_b}}
\newcommand{\mLst}{m_{\Lambda^*}}
\newcommand{\Lb}{\Lambda_b}
\newcommand{\Lst}{\Lambda^*}
\newcommand{\sPlus}{s_+}
\newcommand{\sMinus}{s_-}

\def\eq#1{{Eq.~(\ref{#1})}}
\def\eqs#1#2{{Eqs.~(\ref{#1})--(\ref{#2})}}
\def\fig#1{{Fig.~\ref{#1}}}

\def\Table#1{{Table~\ref{#1}}}

\def\sec#1{{Sect.~\ref{#1}}}

\def\app#1{{Appendix~\ref{#1}}}

\newcommand{\published}[1]{%
\gdef\puB{#1}}
\newcommand{\puB}{}

\allowdisplaybreaks

\begin{document}

\title{\textbf{Heavy quark expansion of $\Lambda_b\to\Lambda^*(1520)$ form factors beyond leading order}}
\author{Marzia Bordone\thanks{marzia.bordone@to.infn.it}}
\affil{Dipartimento di Fisica, Universit\`a di Torino \& INFN, Sezione di Torino, I-10125 Torino, Italy}
\affil{Theoretische Physik 1, Naturwissenschaftlich-Technische Fakult\"at, Universit\"at Siegen, Walter-Flex-Stra{\ss}e 3, D-57068 Siegen, Germany}

 \published{\flushright SI-HEP-2021-04, P3H-21-009\vskip2cm}

\maketitle

\begin{abstract}
I review the parametrisation of the full set of $\Lambda_b\to\Lambda^* (1520)$ form factors in the framework of Heavy Quark Expansion, including next-to-leading-order $\mathcal{O}(\alpha_s)$ and, for the first time, next-to-leading-power $\mathcal{O}(1/m_b)$ corrections. The unknown hadronic parameters are obtained by performing a fit to recent lattice QCD 
 calculations. I investigate the compatibility of the Heavy Quark Expansion and the current lattice data, finding tension between these two approaches in the case of tensor and pseudo-tensor form factors, whose origin could come from an underestimation of the current lattice QCD uncertainties and higher order terms in the Heavy Quark Expansion.
\end{abstract}

\section{Introduction}

The flavour changing neutral current (FCNC)-mediated $b\to s\ell^+\ell^-$ transition plays an important role in the search for physics beyond the Standard Model (SM). Its potential has been extensively studied through the $B\to K^{(*)}\ell^+\ell^-$ decays. Interestingly, the LHCb experiment found some discrepancies with respect to the SM predictions in a few observables: $R_{K}$ and $R_{K^{*}}$, which test universality between the muon and electron final states and the angular coefficient $P_5^\prime$ in the $B\to K^*\mu^+\mu^-$ angular distribution \cite{Aaij:2019wad,Aaij:2017vbb,Aaij:2014ora,Aaij:2020nrf,Aaij:2015oid,Aaij:2013qta}. These hints, together with all available $b\to s\ell^+\ell^-$ data, form a coherent pattern of discrepancies. They can be addressed by introducing New Physics (NP) effects. Low-energy fits point toward a breaking of lepton flavour universality with a combined significance for the NP hypothesis higher than $5\sigma$ \cite{Aebischer:2019mlg,Alguero:2019ptt,Hurth:2020rzx,Ciuchini:2020gvn}. Whether these data show without doubt first signs of NP is not clear yet. Only further measurements with higher statistics or measurements of new processes able to corroborate these data will give a final answer.

A possibility to better understand these data is studying further $b\to s\ell^+\ell^-$-mediated decays, among which baryon decays are promising candidates. The decay channel involving ground-state baryons $\Lb\to\Lambda\mu^+\mu^-$ has already received attention both from the experimental and theoretical point of view. The LHCb experiment measured the $\Lb\to\Lambda\mu^+\mu^-$ angular distribution \cite{Aaij:2015xza,Aaij:2018gwm}, finding good agreement between the measured values of the angular observables and thieir SM predictions \cite{Detmold:2016pkz,Meinel:2016grj,Blake:2019guk} for the angular observables. Even though this result might be discouraging for NP searches, Refs.~\cite{Meinel:2016grj,Blake:2019guk} showed that this is still consistent with the NP hypothesis. In fact, angular distributions describing baryon decays are very different from those in the meson cases, and NP affects them differently. \\
Another possibility is studying excited $\Lambda^*$ states. The LHCb experiment used the decay chain $\Lb\to \Lambda^*(\to pK^-)\ell^+\ell^-$ to measure $R(pK)$, the universality ratio between muons and electrons, finding results consistent with both the SM expectation and the measured values of $R_{K^{(*)}}$ \cite{Aaij:2019bzx}. In this analysis, the various $\Lambda^*$ resonances below a certain mass threshold are not distinguished. However, in Ref.~\cite{Aaij:2015tga} it is shown that the $\Lambda^*(1520)$ is expected to be the most frequent among the $\Lambda^*$ resonances, with quite narrow mass distribution. Therefore, it is motivated to study in detail the $\Lb\to \Lambda^*(1520)\ell^+\ell^-$ decay from both experimental and theoretical point of view, the latter being the main focus of this paper.\\
The determination of the form factors of the $\Lb\to\Lambda^*(1520)$ decays had already been object of study in the literature \cite{Hiller:2007ur,Mott:2011cx,Descotes-Genon:2019dbw,Das:2020cpv,Meinel:2020owd}. The Heavy Quark Expansion (HQE) of the form factors up to next-to-leading (NLO) order in $\alpha_s$ and leading power in $1/m_b$ has been employed \cite{Descotes-Genon:2019dbw,Das:2020cpv}, using Quark Models to constrain the unknown hadronic parameters \cite{Mott:2011cx}. Recently, a lattice QCD determination of the full base of form factors has become available \cite{Meinel:2020owd}, but constrained to the low-recoil region.\\
In this work, I investigate the compatibility between the HQE form factors and the recent lattice QCD determination. At this scope, I perform a HQE of form factors including NLO $\alpha_s$ and next-to-leading power (NLP) $1/m_b$ corrections, the latter being not known so far in the literature. The results are then matched onto the lattice QCD calculation. In the HQE the number of independent, hadronic parameters is reduced compared to the lattice QCD case, introducing strict correlations among the form factors. I perform a fit to the lattice QCD results to obtain the central values, uncertainties, and correlations among the HQE parameters. The comparison between lattice QCD results and HQE predictions shows a tension between the two methods for tensor and pseudo-tensor form factors, whose origin is not yet completely determined.\\
This paper is organised as follows: in \sec{sec:2} I present the HQE of the form factors; in  \sec{sec:3} I discuss the fit to lattice data; in \sec{sec:4} I conclude. \app{app:A} and \app{app:B} report details on the calculation of the form factors and \app{app:C} contains the covariance matrix for the fitted values of the HQE parameters.

\section{Setup}
\label{sec:2}
In the following, I investigate the form factors mediating the transition  $\Lambda_b(p,s_b)\to\Lst(1520)(k,\eta(\lambda_\Lambda),s_\Lambda)$, where $p$ and $k$ are the momenta of the initial and final states, respectively, $s_b$ and $s_\Lambda$ are the rest-frame helicities of the two baryons and $\eta(\lambda_\Lambda)$ is the polarisation vector of the $\Lst$ for each polarisation state $\lambda_\Lambda$. Since in this work I refer to the $\Lambda^*(1520)$, only, I denote this state as $\Lambda^*$ in the following. It is worth noticing that the $\Lambda^*$ is considered stable in this discussion. The subsequent $\Lambda^*$ decay has to be taken into account when comparing to experimental data. In Refs.~\cite{Descotes-Genon:2019dbw,Das:2020cpv} the $\Lambda^*\to N\bar{K}$\footnote{The $\Lambda^*\to N\bar{K}$ is the $\Lambda^*$ decay mode with larger branching fraction and the one that will be employed by the LHCb collaboration to reconstruct the $\Lambda^*$ in future experimental analysis.} decay is discussed, and the four-dimensional differential decay width of the decay chain $\Lambda_b\to\Lambda^*(\to N\bar K)\ell^+\ell^-$ is presented. 

I define the helicity form factors for $\Lambda_b(p,s_b)\to\Lst(k,\eta(\lambda_\Lambda),s_\Lambda)$ as 
\begin{equation}
\label{eq:FFdef}
\begin{aligned}
    \bra{\Lst(k, \eta(\lambda_\Lambda), s_\Lambda)} \bar{s}\gamma^\mu b \ket{\Lb (p, s_b)}
        & = +\bar{u}_\alpha(k, \eta(\lambda_\Lambda), s_\Lambda)
            \left[\sum_{i} F_i(q^2) \Gamma^{\alpha \mu}_{V,i} \right] u(p, s_b)\,,\\
    \bra{\Lst(k, \eta(\lambda_\Lambda), s_\Lambda)} \bar{s}\gamma^\mu \gamma_5 b \ket{\Lb (p, s_b)}
        & = -\bar{u}_\alpha(k, \eta(\lambda_\Lambda), s_\Lambda)
            \left[\sum_{i} G_i(q^2) \gamma_5 \Gamma^{\alpha \mu}_{A,i} \right] u(p, s_b)\,, \\
     \bra{\Lst(k, \eta(\lambda_\Lambda), s_\Lambda)} \bar{s}i\sigma^{\mu\nu} q_\nu b \ket{\Lb (p, s_b)} 
        & = -\bar{u}_\alpha(k, \eta(\lambda_\Lambda), s_\Lambda)
            \left[\sum_{i} T_i(q^2)  \Gamma^{\alpha \mu}_{T,i} \right] u(p, s_b)\,, \\
       \bra{\Lst(k, \eta(\lambda_\Lambda), s_\Lambda)} \bar{s}i\sigma^{\mu\nu} q_\nu\gamma_5 b \ket{\Lb (p, s_b)} 
        & = -\bar{u}_\alpha(k, \eta(\lambda_\Lambda), s_\Lambda)
            \left[\sum_{i} T^5_i(q^2) \gamma_5 \Gamma^{\alpha \mu}_{T5,i} \right] u(p, s_b)\,,      
\end{aligned}
\end{equation}
where $\bar{u}_\alpha$ is the spin $3/2$ projector of a Rarita-Schwinger object \cite{Falk:1991nq}. The Dirac structures $\Gamma^{\alpha\mu}_{L,i}$, with $L=V,A,T,T5$ are given in \app{app:A}. In the cases $L=V,A$ I adopt the parametrisation in Ref.~\cite{Boer:2018vpx}, while for $L=T,T5$ I follow and adapt the parametrisation in Refs.~\cite{Descotes-Genon:2019dbw,Meinel:2020owd}. In the following I use the convention $\sigma^{\mu\nu}= \frac{i}{2}(\gamma^\mu\gamma^\nu-\gamma^\nu\gamma^\mu)$.

\subsection{The Heavy Quark Expansion}
\label{sec:HQE}
In the low-recoil limit, a HQE of the $\Lb\to\Lst$ form factors can be performed. At leading power in $1/m_b$ and leading order in $\alpha_s$, the hadronic matrix element for $\Lb\to \Lst$ transitions reads:
\begin{equation}
\langle \Lambda^*(k,\eta,s_\Lambda)|\bar{s}\Gamma^\mu b| \Lambda_b (p,s_b)\rangle = \sqrt{4} \bar{u}_\alpha(k,\eta, s_\Lambda) \zeta^\alpha(k,\eta, s_\Lambda) \Gamma^\mu u (\mLb v, s_b) \,,
\end{equation}
where $v$ is the velocity of the initial state and $\Gamma^\mu$ denotes a Dirac structure. In the following, I focus on the cases $\Gamma^\mu = \gamma^\mu,\, \gamma^\mu\gamma_5,\, i \sigma^{\mu\nu} q_\nu,\, i \sigma^{\mu\nu} q_\nu\gamma_5$. The most general decomposition for the leading-order and leading-power contribution $\zeta^\alpha$ reads
\begin{equation}
\zeta^\alpha = v^\alpha [\zeta_1+ \zeta_2 \slashed{v}]\,,
\end{equation}
where $\zeta_1$ and $\zeta_2$ are the leading Isgur-Wise (IW) functions. \\
The discussion of $1/m_b$ and $\alpha_s$ corrections closely follows Refs.~\cite{Neubert:1993mb,Grinstein:2004vb,Boer:2014kda}. In this spirit, I replace the (axial-)vector and (pseudo-)tensor currents with:
\begin{align}
\bar{s}\gamma^\mu b \mapsto \bar{s}J^\mu_{V(A)} h_v =\,& (1+C_0^{(v)})\bar{s}\gamma^\mu (\gamma_5) h_v \pm C_1^{(v)} v^\mu \bar{s}(\gamma_5)h_v+ \frac{1}{2 m_b} \bar{s} \Delta J^\mu_{V(A)} h_v  \,, \\
\bar{s}(\gamma_5)i\sigma^{\mu\nu}q_\nu b\mapsto \bar{s} J^\mu_{T(5)} h_v =\,&(1+C_0^{(t)})\bar{s}i\sigma^{\mu\nu}q_\nu h_v + \frac{1}{2 m_b} \bar{s}  \Delta J^\mu_{T(5)} h_v\,. 
\end{align}
The matching coefficients at NLO read \cite{Neubert:1993mb,Grinstein:2004vb}
\begin{equation}
\begin{aligned}
C_0^{(v)}(\mu)=& -\frac{\alpha_s C_F}{4 \pi}\left[3\log\left(\frac{\mu}{m_b}\right)+4\right]+\mathcal{O}(\alpha_s^2)\,, \\
C_1^{(v)}(\mu)=&+\frac{\alpha_s C_F}{2 \pi}+\mathcal{O}(\alpha_s^2)\,, \\
C_0^{(t)}(\mu)=&-\frac{\alpha_s C_F}{4 \pi}\left[5\log\left(\frac{\mu}{m_b}\right)+4\right]+\mathcal{O}(\alpha_s^2)\,.
\end{aligned}
\end{equation}
For numerical purposes, the scale of the Wilson coefficients is set to $\mu\sim 2\, \text{GeV}$. The NLP $1/m_b$ corrections due to the expansion of the current are parametrised as
\begin{equation}
\langle \Lambda^*(k,\eta,s_\Lambda)|\bar{s}\Delta J^\mu_{V(A,T,T5)} b| \Lambda_b (p,s_b)\rangle = \sqrt{4} \sum_i \bar{u}_\alpha (k,\eta, s_\Lambda)\zeta^{\alpha\beta} [\mathcal{O}_{i}^{V(A,T,T5)}]^\mu_\beta u (\mLb v, s_b) \,, 
\end{equation}
where
\begin{equation}
\zeta^{\alpha\beta}  = g^{\alpha\beta} \left[\zeta^\text{SL}_1+\slashed{v}\zeta^\text{SL}_2\right] +v^\alpha v^\beta \left[\zeta^\text{SL}_3+ \slashed{v} \zeta^\text{SL}_4\right]+ v^\alpha \gamma^\beta \left[\zeta_5^\text{SL}+\slashed{v}\zeta^\text{SL}_6\right]\,.
\label{eq:subleading_expansion}
\end{equation}
The functions $\zeta^\text{SL}_{1\dots6}$ are the subleading Isgur-Wise functions, and they correspond to all the independent Dirac structures that can appear in $ \zeta^{\alpha\beta}$. The possible operators $\mathcal{O}^{\mu\beta}_{i}$ are listed in Ref.~\cite{Neubert:1993mb}. Out of the possible six of them, only the operator $[\mathcal{O}_1^\Gamma]^\mu_\beta$ arises at order $1/m_b$,  while the others are suppressed by $\mathcal{O}(\alpha_s/m_b)$ and therefore are beyond the precision here required. Therefore, the only contributions that I consider for this analysis come from:
\begin{equation}
\begin{aligned}
\left[ \mathcal{O}_1^V \right]^\mu_\beta=& +\gamma^\mu \gamma_\beta \,, &  \left[\mathcal{O}_1^A\right]^\mu_\beta=& -\gamma_5 \gamma^\mu \gamma_\beta\,,\\
\left[\mathcal{O}_1^T\right]^\mu_\beta=& +i\sigma^{\mu\nu}q_\nu \gamma_\beta \,, & \left[\mathcal{O}_1^{T5}\right]^\mu_\beta=& +i\gamma_5\sigma^{\mu\nu}q_\nu \gamma_\beta \,,
\end{aligned}
\label{eq:subleading_op}
\end{equation}
By means of Dirac algebra, and using the properties of Rarita-Schwinger objects in Ref.~\cite{Falk:1991nq}, inserting \eq{eq:subleading_op} in \eq{eq:subleading_expansion} yields
\begin{align}
\langle \Lambda^*(k,\eta,s_\Lambda)|\bar{s}\Delta J^\mu_{V} b| \Lambda_b (p,s_b)\rangle  = 2 \bigg\lbrace&2\bar{u}_\alpha(k,\eta, s_\Lambda)  u (\mLb v, s_b)  g^{\alpha\mu}(\zeta^\text{SL}_1+\zeta^\text{SL}_2) \nonumber\\
&+\bar{u}_\alpha (k,\eta, s_\Lambda) \gamma^\mu (\mLb v, s_b) v^\alpha (\zeta^\text{SL}_3-\zeta^\text{SL}_4-2 \zeta^\text{SL}_2-2 \zeta^\text{SL}_5) \nonumber\\
&+2 \bar{u}_\alpha (k,\eta, s_\Lambda) u (\mLb v, s_b) v^\alpha v^\mu (\zeta^\text{SL}_4+2\zeta^\text{SL}_6) \bigg\rbrace \,,\\
\langle \Lambda^*(k,\eta,s_\Lambda)|\bar{s}\Delta J^\mu_{A} b| \Lambda_b (p,s_b)\rangle  = 2 \bigg\lbrace&2\bar{u}_\alpha (k,\eta, s_\Lambda)\gamma_5 u (\mLb v, s_b)  g^{\alpha\mu}(-\zeta^\text{SL}_1+\zeta^\text{SL}_2) \nonumber\\
&-\bar{u}_\alpha (k,\eta, s_\Lambda) \gamma_5\gamma^\mu (\mLb v, s_b) v^\alpha (\zeta^\text{SL}_3+\zeta^\text{SL}_4+2 \zeta^\text{SL}_2+2 \zeta^\text{SL}_5 ) \nonumber\\
&+2 \bar{u}_\alpha (k,\eta, s_\Lambda)\gamma_5 u (\mLb v, s_b) v^\alpha v^\mu (\zeta^\text{SL}_4-2\zeta^\text{SL}_6)\bigg\rbrace \,,\\
\langle \Lambda^*(k,\eta,s_\Lambda)|\bar{s}\Delta J^\mu_{T} b| \Lambda_b (p,s_b)\rangle  = 2 \bigg\lbrace&-\bar{u}_\alpha (k,\eta, s_\Lambda)\gamma^\mu u (\mLb v, s_b) v^\alpha\bigg[2\mLb\zeta_1^\text{SL}+2\mLst\zeta_2^\text{SL} \nonumber \\
&+(\mLb+\mLst)(\zeta_3^\text{SL}+2\zeta_6^\text{SL})+\frac{\mLst^2+\mLb \mLst-q^2}{\mLb}\zeta_4^\text{SL}\bigg]\nonumber \\
&+2\bar{u}_\mu (k,\eta, s_\Lambda) u (\mLb v, s_b) \bigg[(\mLb-\mLst) \zeta_1^\text{SL}-\frac{\mLst^2-\mLst\mLb-q^2}{\mLb}\zeta_2^\text{SL}\bigg]  \nonumber \\
&+2\bar{u}_\alpha(k,\eta, s_\Lambda)\gamma^\mu  u (\mLb v, s_b) k^\alpha(\zeta_1^\text{SL}-\zeta_2^\text{SL})  \nonumber\\ 
&+\bar{u}_\alpha (k,\eta, s_\Lambda) u (\mLb v, s_b) v^\alpha k^\mu(\zeta_3^\text{SL}+2\zeta_2^\text{SL}+\zeta_4^\text{SL}+2\zeta_6^\text{SL}) \nonumber\\
&+4\bar{u}_\alpha (k,\eta, s_\Lambda) u (\mLb v, s_b) v^\mu k^\alpha\zeta_2^\text{SL} \nonumber\\
&+\bar{u}_\alpha (k,\eta, s_\Lambda) u (\mLb v, s_b) v^\alpha v^\mu \nonumber\\
&\times \big[\mLb\zeta_3^\text{SL}-2\mLb\zeta_2^\text{SL}+(2\mLst-\mLb)\zeta_4^\text{SL}+2\mLb \zeta_6^\text{SL}\big]\bigg\rbrace\,, \\
\langle \Lambda^*(k,\eta,s_\Lambda)|\bar{s}\Delta J^\mu_{T5} b| \Lambda_b (p,s_b)\rangle  = 2  \bigg\lbrace&-\bar{u}_\alpha (k,\eta, s_\Lambda)\gamma_5\gamma^\mu u (\mLb v, s_b) v^\alpha\bigg[\mLb\zeta_1^\text{SL}+2\mLst\zeta_2^\text{SL} \nonumber \\
&+(\mLb-\mLst)(\zeta_3^\text{SL}+2\zeta_6^\text{SL})-\frac{\mLst^2-\mLb \mLst-q^2}{\mLb}\zeta_4^\text{SL}\bigg]\nonumber \\
&+2 \bar{u}_\mu (k,\eta, s_\Lambda) \gamma_5u (\mLb v, s_b) \bigg[(\mLb+\mLst)\zeta_1^\text{SL}+\frac{\mLst^2+\mLst\mLb-q^2}{\mLb}\zeta_2^\text{SL}\bigg] \nonumber\\
&+2\bar{u}_\alpha(k,\eta, s_\Lambda)\gamma_5\gamma^\mu  u (\mLb v, s_b) k^\alpha(\zeta_1^\text{SL}+\zeta_2^\text{SL})  \nonumber\\ 
&+\bar{u}_\alpha (k,\eta, s_\Lambda) \gamma_5 u (\mLb v, s_b) v^\alpha k^\mu(+\zeta_3^\text{SL}-2\zeta_2^\text{SL}-\zeta_4^\text{SL}+2\zeta_6^\text{SL}) \nonumber\\
&-4\bar{u}_\alpha (k,\eta, s_\Lambda) \gamma_5u (\mLb v, s_b) v^\mu k^\alpha \zeta_2^\text{SL}  \nonumber\\
&+\bar{u}_\alpha (k,\eta, s_\Lambda)\gamma_5 u (\mLb v, s_b) v^\alpha v^\mu \nonumber\\
&\times \big[\mLb\zeta_3^\text{SL}+2\mLb\zeta_2^\text{SL}+(2\mLst+\mLb)\zeta_4^\text{SL}+2\mLb \zeta_6^\text{SL}\big]\bigg\rbrace\,.
\end{align}
The number of independent subleading IW functions can be reduced by using equations of motions. In particular, for this decay, the relation $v_\beta \zeta^{\alpha\beta}=0$ gives the following conditions:
\begin{align}
\zeta_1^\text{SL}+\zeta_3^\text{SL}+\zeta_6^\text{SL} =0\,, \\
\zeta_2^\text{SL}+\zeta_4^\text{SL}+\zeta_5^\text{SL} =0\,. 
\end{align}
I choose to retain as independent quantities the subleading IW functions $\zeta_1^\text{SL}$, $\zeta_2^\text{SL}$, $\zeta_3^\text{SL}$  and $\zeta_4^\text{SL}$.\\
Corrections to the form factors arise also by of non-local insertions of the heavy quark Lagrangian at order $1/m_b$. Following the discussions in Refs.\cite{Falk:1992ws,Falk:1992wt,Neubert:1993mb}, non-local insertion of the kinetic operator yields a universal shift proportional to the tree-level matrix elements. Hence I reabsorb such shift in a redefinition of the leading order IW function $\zeta_1$. Non-local insertion of the chromomagnetic operator can be parametrised as \cite{Leibovich:1997az}
\begin{equation}
R_{\mu\nu}^\alpha \bar{u}_\alpha (k,\eta, s_\Lambda) \Gamma \frac{1+\slashed{v}}{2} i\sigma^{\mu\nu}u (\mLb v, s_b)\,,
\end{equation}
where the object $R_{\mu\nu}^\alpha$ is an antisymmetric tensor in the indices $\mu$ and $\nu$ and contains the velocity $v$. $\Gamma$ symbolises all the possible Dirac structures which mediate  $\Lambda_b\to\Lambda^*$ transitions. Using equations of motion, it can be shown that no possible form of $R_{\mu\nu}^\alpha$ gives a non-zero contribution for the chromomagnetic operator.\\

The expressions of the form factors in terms of the leading and subleading IW functions are obtained by matching the helicity amplitudes with their HQE expansion. For the vector current, by comparing \eqs{eq:HAVQCD:1}{eq:HAVQCD:4} to \eqs{eq:HAVHQE:1}{eq:HAVHQE:4}, I get
\begin{align}
F_{1/2,0} = &\frac{\sqrt{s_+}}{2\mLb^{3/2}\mLst^{3/2}} \bigg\{\zeta_1 s_-\left[(1+C_0^{(v)})+ C_1^{(v)}\frac{s_+}{2\mLb(\mLb+\mLst)}\right]  \nonumber\\
		+& \zeta_2 s_-\left[(1+C_0^{(v)})\frac{\mLst^2+\mLst\mLb-q^2}{\mLb(\mLb+\mLst)}+ C_1^{(v)}\frac{s_+}{2\mLb(\mLb+\mLst)}\right]  \nonumber\\
		-& \zeta_1^\text{SL} \frac{\lambda+\mLb^2(\mLst^2-\mLb^2+q^2)}{\mLb^2(\mLst+\mLb)}-\zeta_2^\text{SL} \frac{(\mLst^2-\mLb^2+q^2)}{(\mLst+\mLb)} \nonumber\\
		-&s_-\left[ \zeta_3^\text{SL} \frac{(2\mLst^2+3\mLst\mLb+\mLb^2-2 q^2)}{2\mLb^{2}(\mLst+\mLb)}-\zeta_4^\text{SL} \frac{(\mLst^2+3\mLst\mLb+2\mLb^2-q^2)}{2\mLb^{2}(\mLst+\mLb)}\right]\bigg\}\,, \label{eq:F120}\\
F_{1/2,t} = &\frac{\sqrt{s_-} \sPlus}{2\mLb^{3/2}\mLst^{3/2}}\bigg\{\zeta_1 \left[(1+C_0^{(v)})- C_1^{(v)}\frac{\mLst^2-\mLb^2-q^2}{2\mLb(\mLb-\mLst)}\right]  \nonumber\\    
	      -& \zeta_2 \frac{1}{2\mLb(\mLb-\mLst)}\left[2(1+C_0^{(v)})(\mLst^2-\mLst\mLb-q^2)+ C_1^{(v)}(\mLst^2-\mLb^2-q^2)\right]  \nonumber\\
	      -&\frac{1}{\mLb-\mLst} \left[\frac{s-\mLst^2}{\mLb^2}\zeta_1^\text{SL}-\zeta_2^\text{SL}\right] \nonumber\\
	      +&\left[\frac{2\mLst^2-\mLst\mLb-\mLb^2-2q^2}{2\mLb^2(\mLb-\mLst)}\zeta_3^\text{SL}-\frac{\mLst^2+\mLst\mLb-2\mLb^2-q^2}{2\mLb^2(\mLb-\mLst)}\zeta_4^\text{SL}\right]\,, \\
F_{1/2,\perp} = &\frac{\sqrt{s_+}}{2\mLb^{3/2}\mLst^{3/2}}\bigg\{
			 s_-(1+ C_0^{(v)})(\zeta_1-\zeta_2)-\mLst (\zeta_1^\text{SL}+\zeta_2^\text{SL}) +\frac{s_-}{2\mLb}   (\zeta_3^\text{SL}+\zeta_4^\text{SL})\bigg\}\,, \\
F_{3/2,\perp} =  &-\frac{\sqrt{s_+}}{2\mLb^{3/2}\mLst^{1/2}}\bigg\{\zeta_1^\text{SL}+\zeta_2^\text{SL}\bigg\}\,,
\end{align}
and for the axial vector current, by matching \eqs{eq:HAAQCD:1}{eq:HAAQCD:4} to \eqs{eq:HAAHQE:1}{eq:HAAHQE:4} I obtain
\begin{align}
G_{1/2,0} = &\frac{\sqrt{s_-}}{2\mLb^{3/2}\mLst^{3/2}} \bigg\{\zeta_1 s_+\left[(1+C_0^{(v)})+C_1^{(v)}\frac{s_-}{2\mLb(\mLb-\mLst)}\right]  \nonumber\\
		-& \zeta_2 s_+\left[(1+C_0^{(v)})\frac{\mLst^2-\mLst\mLb-q^2}{\mLb(\mLb-\mLst)}- C_1^{(v)}\frac{s_-}{2\mLb(\mLb-\mLst)}\right]  \nonumber\\
		-& \zeta_1^\text{SL} \frac{\lambda+\mLb^2(\mLst^2-\mLb^2+q^2)}{\mLb^2(\mLb-\mLst)}+\zeta_2^\text{SL} \frac{(\mLst^2-\mLb^2+q^2)}{(\mLb-\mLst)} \nonumber\\
		-&s_+\left[ \zeta_3^\text{SL} \frac{(2\mLst^2-3\mLst\mLb+\mLb^2-2 q^2)}{2\mLb^{2}(\mLb-\mLst)}+\zeta_4^\text{SL} \frac{(\mLst^2-3\mLst\mLb+2\mLb^2-q^2)}{2\mLb^{2}(\mLb-\mLst)}\right]\bigg\}\,,\\
G_{1/2,t} = &\frac{\sqrt{s_+} \sMinus}{2\mLb^{3/2}\mLst^{3/2}}\bigg\{\zeta_1 \left[(1+C_0^{(v)})- C_1^{(v)}\frac{\mLst^2-\mLb^2-q^2}{2\mLb(\mLb+\mLst)}\right]  \nonumber\\    
	      +& \zeta_2 \frac{1}{2\mLb(\mLb+\mLst)}\left[2(1+C_0^{(v)})(\mLst^2+\mLst\mLb-q^2)- C_1^{(v)}(\mLst^2-\mLb^2-q^2)\right]  \nonumber\\
	      -&\frac{1}{\mLb+\mLst} \left[\frac{s-\mLst^2}{\mLb^2}\zeta_1^\text{SL}+\zeta_2^\text{SL}\right] \nonumber\\
	      +&\left[\frac{2\mLst^2+\mLst\mLb-\mLb^2-2q^2}{2\mLb^2(\mLb+\mLst)}\zeta_3^\text{SL}+\frac{\mLst^2-\mLst\mLb-2\mLb^2-q^2}{2\mLb^2(\mLb+\mLst)}\zeta_4^\text{SL}\right]\,, \label{eq:FF_HQE_G12t} \\
G_{1/2,\perp} = &\frac{\sqrt{s_-}}{2\mLb^{3/2}\mLst^{3/2}}\bigg\{
			 s_+\left[(1+ C_0^{(v)})(\zeta_1+\zeta_2)\right] +\mLst (\zeta_1^\text{SL}-\zeta_2^\text{SL})+\frac{s_+}{2\mLb}   (\zeta_3^\text{SL}-\zeta_4^\text{SL})\bigg\}\,, \\
G_{3/2,\perp} =  &-\frac{\sqrt{s_-}}{2\mLb^{3/2}\mLst^{1/2}}\bigg\{\zeta_1^\text{SL}-\zeta_2^\text{SL}\bigg\}\,.
\end{align}
For the tensor current, the comparison between \eqs{eq:HATQCD:1}{eq:HATQCD:3} and \eqs{eq:HATHQE:1}{eq:HATHQE:3} yields
\begin{align}
T_{1/2,0} = & \frac{\sqrt{\sPlus}}{\mLb^{3/2}\mLst^{1/2}}\bigg\{(1+C_0^{(t)}) (\zeta_1-\zeta_2) \sMinus-\frac{\mLst^2+\mLb^2-q^2}{\mLb}(\zeta_1^\text{SL}-\zeta_2^\text{SL})\nonumber\\
		&-\frac{\sMinus}{2\mLb}(\zeta_4^\text{SL}+\zeta_3^\text{SL})\bigg\}\,, \\
T_{1/2,\perp} = &  \frac{\sqrt{\sPlus}}{\mLb^{3/2}\mLst^{1/2}}\bigg\{ (1+C_0^{(t)}) \left[\zeta_1+\frac{\mLst(\mLb+\mLst)-q^2}{\mLb(\mLst+\mLb)}\zeta_2\right]\sMinus \nonumber \\
			+& \mLst\frac{\mLst(\mLb-\mLst)+q^2}{\mLb(\mLst+\mLb)}\zeta_1^\text{SL} + \frac{\mLst(\mLb-\mLst)}{\mLst+\mLb}\zeta_2^\text{SL}  \nonumber\\
			+&\frac{\sMinus}{2\mLb}\left[ -\zeta_3^\text{SL} +\frac{\mLst(\mLb+\mLst)-q^2}{\mLb(\mLst+\mLb)}\zeta_4^\text{SL}\right]\bigg\} \,,\\
T_{3/2,\perp} = &  +\frac{\sqrt{\sPlus}}{\mLb^{3/2}\mLst^{1/2}}\bigg\{-\zeta_1^\text{SL}(\mLb-\mLst)+\frac{\mLst^2-\mLb \mLst-q^2}{\mLb}\zeta_2^\text{SL}\bigg\}\,,
\end{align}
while for the axial-tensor form factors the comparison between \eqs{eq:HAT5QCD:1}{eq:HAT5QCD:3} and \eqs{eq:HAT5HQE:1}{eq:HAT5HQE:3} gives
\begin{align}
T_{1/2,0}^5 = & \frac{\sqrt{\sMinus}}{\mLb^{3/2}\mLst^{1/2}}\bigg\{(1+C_0^{(t)}) (\zeta_2+\zeta_1) \sPlus-\frac{\mLst^2+\mLb^2-q^2}{\mLb}(\zeta_1^\text{SL}-\zeta_2^\text{SL})\nonumber\\
		-&\frac{\sPlus}{2\mLb}(\zeta_3^\text{SL}-\zeta_4^\text{SL})\bigg\}\,, \\
T_{1/2,\perp}^5 = &  \frac{\sqrt{\sMinus}}{\mLb^{3/2}\mLst^{1/2}}\bigg\{ (1+C_0^{(t)}) \left[\zeta_1-\frac{\mLst(-\mLb+\mLst)-q^2}{\mLb(\mLb-\mLst)}\zeta_2\right]\sPlus \nonumber \\
			+&\mLst \frac{\mLst(+\mLb+\mLst)-q^2}{\mLb(\mLb-\mLst)}\zeta_1^\text{SL} + \frac{\mLst(\mLb+\mLst)}{\mLb-\mLst}\zeta_2^\text{SL}  \nonumber\\
			-&\frac{\sPlus}{2\mLb}\left[ \zeta_3^\text{SL} +\frac{\mLst(-\mLb+\mLst)-q^2}{\mLb(\mLb-\mLst)}\zeta_4^\text{SL}\right]\bigg\} \,,\\
T_{3/2,\perp}^5 = &  -\frac{\sqrt{\sMinus}}{\mLb^{3/2}\mLst^{1/2}}\bigg\{\zeta_1^\text{SL}(\mLb+\mLst)+\frac{\mLst^2+\mLb \mLst-q^2}{\mLb}\zeta_2^\text{SL}\bigg\}\,. \label{eq:T532perp}
\end{align}
The expressions in \eqs{eq:F120}{eq:T532perp} have been checked against the results in Ref.~\cite{Das:2020cpv}, where the HQE for the form factors including NLO $\alpha_s$ corrections is presented. I find agreement with the expressions reported in Ref.~\cite{Das:2020cpv}, apart from a sign flip in the term proportional to $C_1^{(v)}\zeta_2$ in $G_{1/2,t}$. This misalignment is not expected to invalidate the analysis in Ref.~\cite{Das:2020cpv} since it appears in a term that is $\alpha_s$ suppressed. In the following analysis, I stick to my findings and adopt the signs in \eq{eq:FF_HQE_G12t}.

\section{Form Factors relations and comparison with lattice QCD results}
\label{sec:3}
The first lattice QCD calculation for the full basis of $\Lb\to\Lst$ form factors is presented in Ref.~\cite{Meinel:2020owd}. The calculation is performed in the low-recoil region, very close to the zero-recoil point $q^2_\text{max} =(\mLb-\mLst)^2$. Two lattice points per form factor are obtained, allowing to determine the normalisation and the slope of each form factor. In the kinematical limit where the lattice QCD computation is valid, it is more convenient to substitute the variable $q^2$ with the adimensional variable $w= p\cdot k/(2\mLb\mLst) = (\mLb^2+\mLst^2-q^2)/(2\mLb\mLst)$, where the zero-recoil point corresponds to $w=1$. The continuum extrapolation of the results in Ref.~\cite{Meinel:2020owd} is performed using the following functional form for each of the form factor $f_i$: 
\begin{equation}
f_i = F_i+A_i(w-1)\,.
\label{eq:latticeFFs}
\end{equation}
Values for the coefficients $F_i$ and $A_i$ and their covariance matrix are given in ancillary files of Ref.~\cite{Meinel:2020owd}.  \\
Given that both the results in Ref.~\cite{Meinel:2020owd} and the parametrisation based on HQE in \sec{sec:2} are valid in the low-recoil region, the former is suitable to extract the unknown, hadronic parameters for the leading and sub-leading IW functions introduced in \sec{sec:2}. I want to stress that it is not possible to extrapolate these results to the large-recoil region without any further information on the form factors valid at low $q^2$.\\ The form factor base employed in Ref.~\cite{Meinel:2020owd} differs from the one presented in \sec{sec:2} and the matching between the two is given in \app{app:B}. In the following, I denote with capital letters the HQE base and with lower cases the base of Ref.~\cite{Meinel:2020owd}.
 
\subsection{Relations in the zero-recoil point}
I study the form factors first at the zero-recoil point. At this particular kinematical configuration, all the axial-vector and pseudo-tensor HQE form factors become zero because they are weighted by the factor $s_-$. The remainder is further simplified since the terms associated with  $\zeta_1, \zeta_2, \zeta_3^\text{SL}, \zeta_{4}^\text{SL}$, are always proportional to $s_-$, hence vanish, leaving only $\zeta_1^\text{SL}$ and $\zeta_{2}^\text{SL}$ to determine the form factors in the zero-recoil point. Even more interestingly, from \eqs{eq:F120}{eq:T532perp} it can be seen that only the combination $\zeta_1^\text{SL}+ \zeta_{2}^\text{SL}$ appears, with different weights for each of the form factor. Therefore, HQE provides predictions for ratios of form factors at zero-recoil independent of the IW functions. They read:
\begin{equation}
\begin{aligned}
\label{eq:ZR_HQE}
\frac{F_{1/2,0}}{F_{1/2,\perp}} =&\,  \frac{F_{1/2,0}}{F_{3/2,\perp}} = -2\frac{\mLb-\mLst}{\mLb+\mLst} = -1.15  \,,  &    \frac{F_{1/2,\perp}}{F_{3/2,\perp}} =&\,1  \,,  \\
\frac{T_{1/2,0}}{T_{1/2,\perp}} =&-2\frac{\mLb+\mLst}{\mLb-\mLst}=-3.48 \,,  & \frac{T_{1/2,0}}{T_{3/2,\perp}} =&\frac{2\mLst}{\mLb-\mLst}=0.74  \,,  \\
  \frac{T_{1/2,\perp}}{T_{3/2,\perp}} =&-\frac{\mLst}{\mLb+\mLst}=-0.21   \,,
\end{aligned}
\end{equation}
where errors on the baryon masses have been neglected.
These predictions have to be compared with the results in Ref.~\cite{Meinel:2020owd}. Extracting pseudodata for the parameters entering the lattice QCD form factors, according to their central values and covariance matrix, I find (in the HQE basis)
\begin{equation}
\begin{aligned}
\label{eq:ZR_lattice}
\frac{F_{1/2,0}^\text{latt}}{F_{1/2,\perp}^\text{latt}} =&-0.63\pm 1.82 \,,  & \frac{F_{1/2,0}^\text{latt}}{F_{3/2,\perp}^\text{latt}} =&-0.94\pm0.14  \,,  &  \frac{F_{1/2,\perp}^\text{latt}}{F_{3/2,\perp}^\text{latt}} =&\,1.48\pm 0.39  \,,  \\
\frac{T_{1/2,0}^\text{latt}}{T_{1/2,\perp}^\text{latt}} =&-5.62\pm 7.86 \,,  & \frac{T_{1/2,0}^\text{latt}}{T_{3/2,\perp}^\text{latt}} =&+1.03\pm0.20  \,,  &  \frac{T_{1/2,\perp}^\text{latt}}{T_{3/2,\perp}^\text{latt}} =&-0.18\pm0.05  \,.
\end{aligned}
\end{equation}
The central values showed above are the medians of the distributions and the errors correspond to the $68\%$ intervals. Concerning the ratios $F_{1/2,0}^\text{latt}/F_{1/2,\perp}^\text{latt}$ and $T_{1/2,0}^\text{latt}/T_{1/2,\perp}^\text{latt}$, I stress that their uncertainties are very large to reflect the fact that their distributions are highly non gaussian. For the other ratios, the gaussian approximation well describes lattice data. The predictions for the ratios obtained in the HQE framework and using the lattice QCD results agree within $\sim1.1\sigma$.\\

\subsection{Form factors parametrisation and fit}
\label{sec:3.1}
The HQE of the $\Lb\to\Lst$ form factors presented in \sec{sec:2} depends on 6 unknown IW functions. The HQE does not predict the $w$ dependence of the IW functions, and it has to be inferred from other principles. Since the formalism of HQE is valid mainly in the low-recoil region for $b\to s$ transitions, the form factors can be expanded around the zero-recoil point. Substituting $q^2$ with $w$, the IW functions $\zeta_i $ can be expanded as 
\begin{equation}
\zeta_i = \sum_{n=0}^N \frac{\zeta_i^{(n)}}{n!} (w-1)^n\,.
\label{eq:IWfunct_expansion}
\end{equation}
The truncation order $N$ of the expansion depends on the precision required and how far from $w=1$ the form factors are evaluated. The parameters $\zeta_i^{(n)}$ are unknown and have to be fixed using some dynamical information. Notice that since the $\Lst$ is not a ground state baryon, the HQE does not predict any normalisation for the leading IW functions. \\
For convenience, I perform a fit to the lattice QCD data using the base in Ref.~\cite{Meinel:2020owd}, using the chiral and continuum extrapolation there provided to obtain pseudo-points for the form factors. Since the lattice QCD data do not provide information on the curvature of the form factors, it is useful to express the HQE form factors as in the form of \eq{eq:latticeFFs}. At this scope, I use the parametrisation in \eq{eq:IWfunct_expansion} up to $N=1$ and then re-expand the full form factors up to the first order in $w-1$. 
After this procedure, it can be noticed that \textit{i)} the parameters $\zeta_1^{(1)}$ and  $\zeta_2^{(1)}$ appear always in the combination $\zeta_1^{(1)}+\zeta_2^{(1)}$ and \textit{ii)} the parameters $\zeta_3^{\text{SL},(1)}$ and $\zeta_4^{\text{SL},(1)}$ appear always in the combination $\zeta_4^{\text{SL},(1)}-\zeta_3^{\text{SL},(1)}$. Therefore these parameters cannot be determined on their own, but only the combinations $\zeta_1^{(1)}+\zeta_2^{(1)}$ and $\zeta_4^{\text{SL},(1)}-\zeta_3^{\text{SL},(1)}$ are determined. This makes the number of independent, unknown HQE parameters $10$. \\
Before discussing the fit results, a couple of technical comments are in order:
\begin{enumerate}
\item
Given the available information in Ref.~\cite{Meinel:2020owd}, it is possible to use two pseudo-points for each form factor. I choose to evaluate them at $w=1.02$ and $w=1.04$. Given that the HQE parametrisation depends on fewer parameters than the lattice QCD one, it is possible to perform a fit to only a subset of the lattice QCD data. In particular, I choose to employ the data on the vector and axial-vector form factors and provide predictions for the tensor and pseudo-tensor form factors based on the fit results. I comment on the consequences of this choice in the following. I stress that this is a common procedure and has been already employed in Ref.~\cite{Bernlochner:2018kxh}.
\item
The HQE based form factors are affected by uncertainties due to the unknown contributions from higher orders of expansion. By naive dimensional arguments these contributions are expected to be roughly $\mathcal{O}(\text{few}\,\%)$. Hence, I introduce an uncorrelated $1\%$ uncertainty on all HQE expressions of the form factors to take these effects into account. Comments on this choice can be found later in the text.
\end{enumerate}

\begin{table}
\begin{center}
\renewcommand{\arraystretch}{1.2} 
\begin{tabular}{ c c }
\toprule
Parameter & Best fit point \\
\midrule
$\zeta_1^{(0)}$ & $0.454 \pm 0.070$ \\
$\zeta_2^{(0)}$ &  $-0.0303 \pm 0.0552$\\
$\zeta_1^{(1)}+\zeta_2^{(1)}$  &  $0.113 \pm 0.024$\\
$\zeta_1^{\text{SL},(0)}$  &  $0.125\pm0.038$ \\
$\zeta_1^{\text{SL},(1)}$  &  $0.0487 \pm 0.0614$\\
$\zeta_2^{\text{SL},(0)}$  &  $0.0110\pm0.0363 $\\
$\zeta_2^{\text{SL},(1)}$  &  $0.00362 \pm 0.06184$  \\
$\zeta_3^{\text{SL},(0)}$  &  $0.228\pm0.190$\\
$\zeta_4^{\text{SL},(0)}$  &  $0.0883\pm0.185$\\
$\zeta_4^{\text{SL},(1)}-\zeta_3^{\text{SL},(1)}$  & $-0.0267\pm0.0773$ \\
\toprule
\end{tabular}
\caption{Best fit points for the HQE parameters.}
\label{tab:fit_values}
\end{center}
\end{table}

The fit is performed with a $\chi^2$ minimisation, yielding at the minimum $\chi^2_\text{min}/\text{d.o.f.} \sim 0.25$. This low value is a direct consequence of the poor current knowledge of the exact size of the theory uncertainties and their correlations. If the theory uncertainties were uncorrelated, the fit procedure would indicate that their natural size is smaller than the one inferred from HQE. However, such a low value for the $\chi^2_\text{min}/\text{d.o.f.}$ could also indicate that strong correlations among unknown higher-order terms in the HQE have been neglected. At the current status, it is not possible to obtain a more precise estimation of theory uncertainties and their correlations. Therefore, I retain the conservative choice of an uncorrelated $1\%$ uncertainty on all the form factors in the analysis.\\

\begin{figure}[H]
\begin{center}
  \begin{tabular}{cc}
       \includegraphics[scale=0.32]{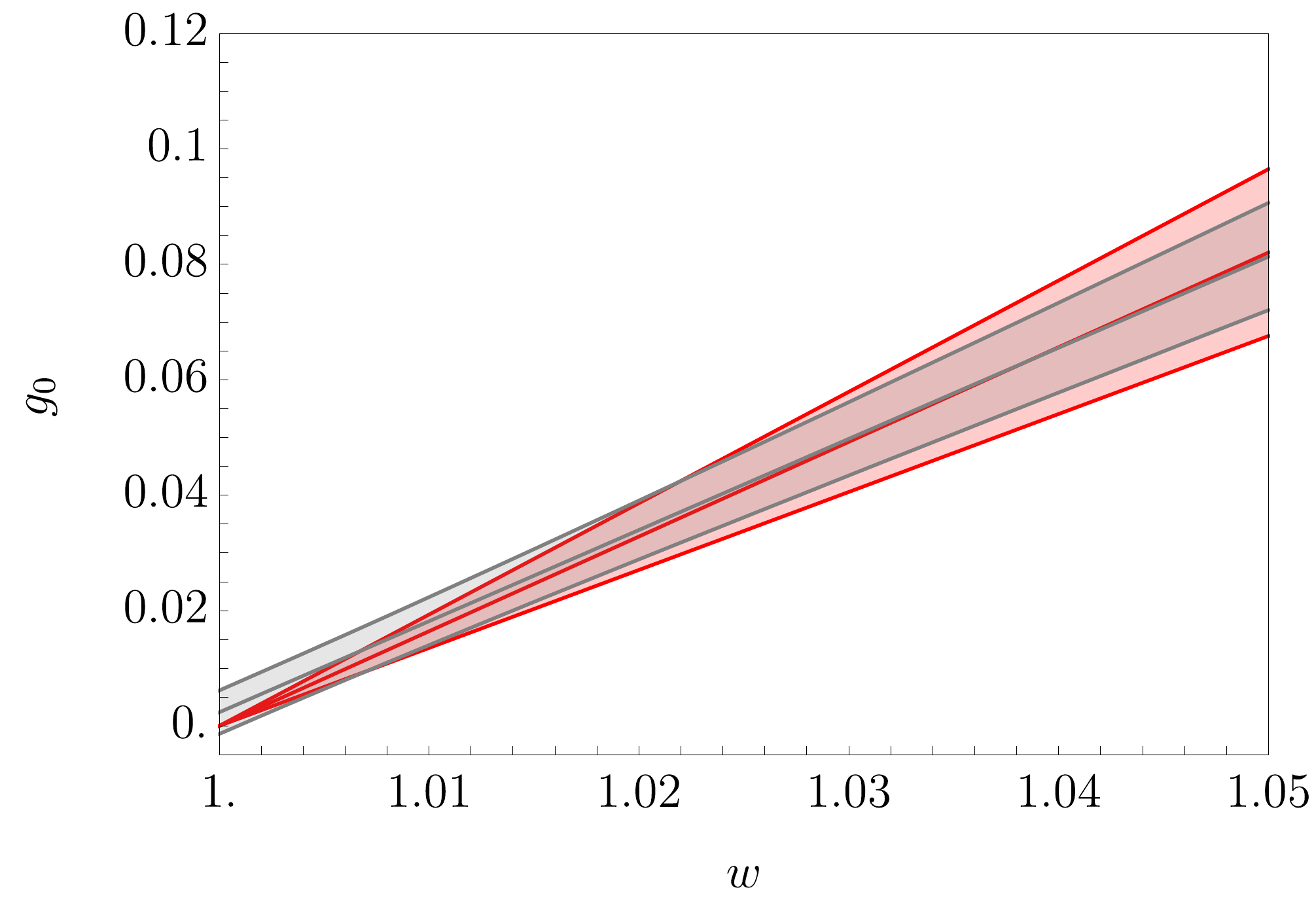}  &
       \includegraphics[scale=0.32]{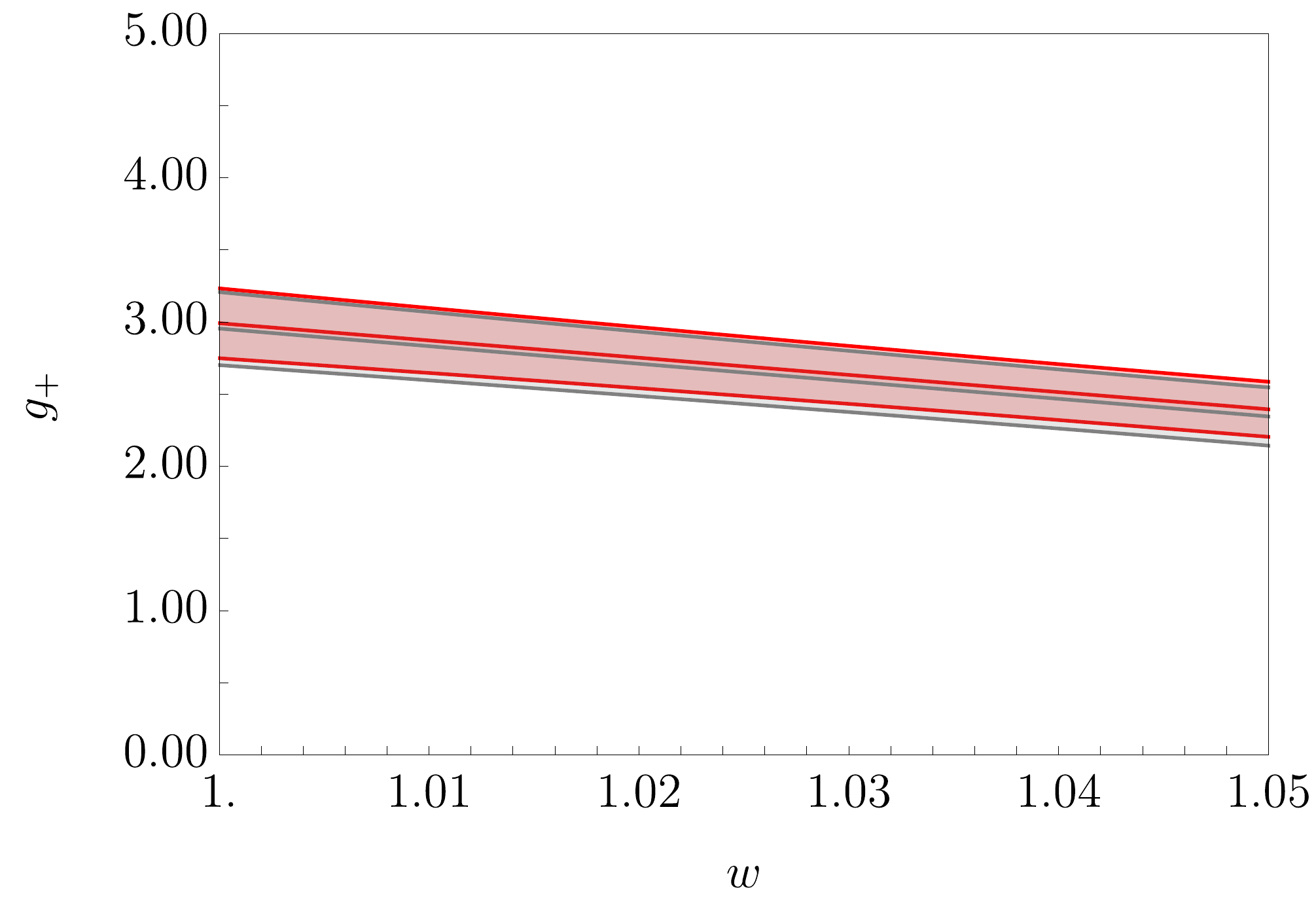}  \\[1.25em]
        \includegraphics[scale=0.32]{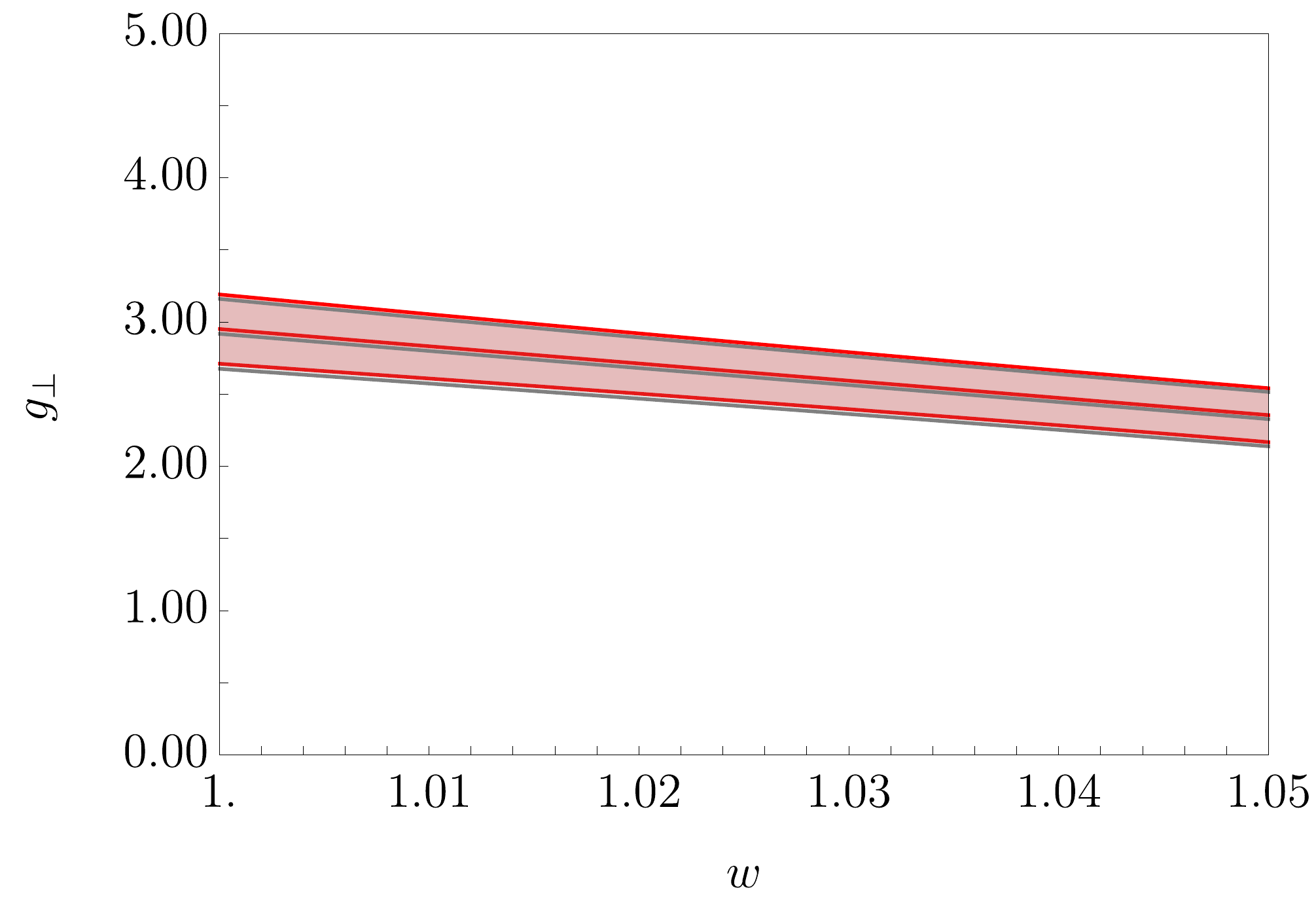} &
        \includegraphics[scale=0.32]{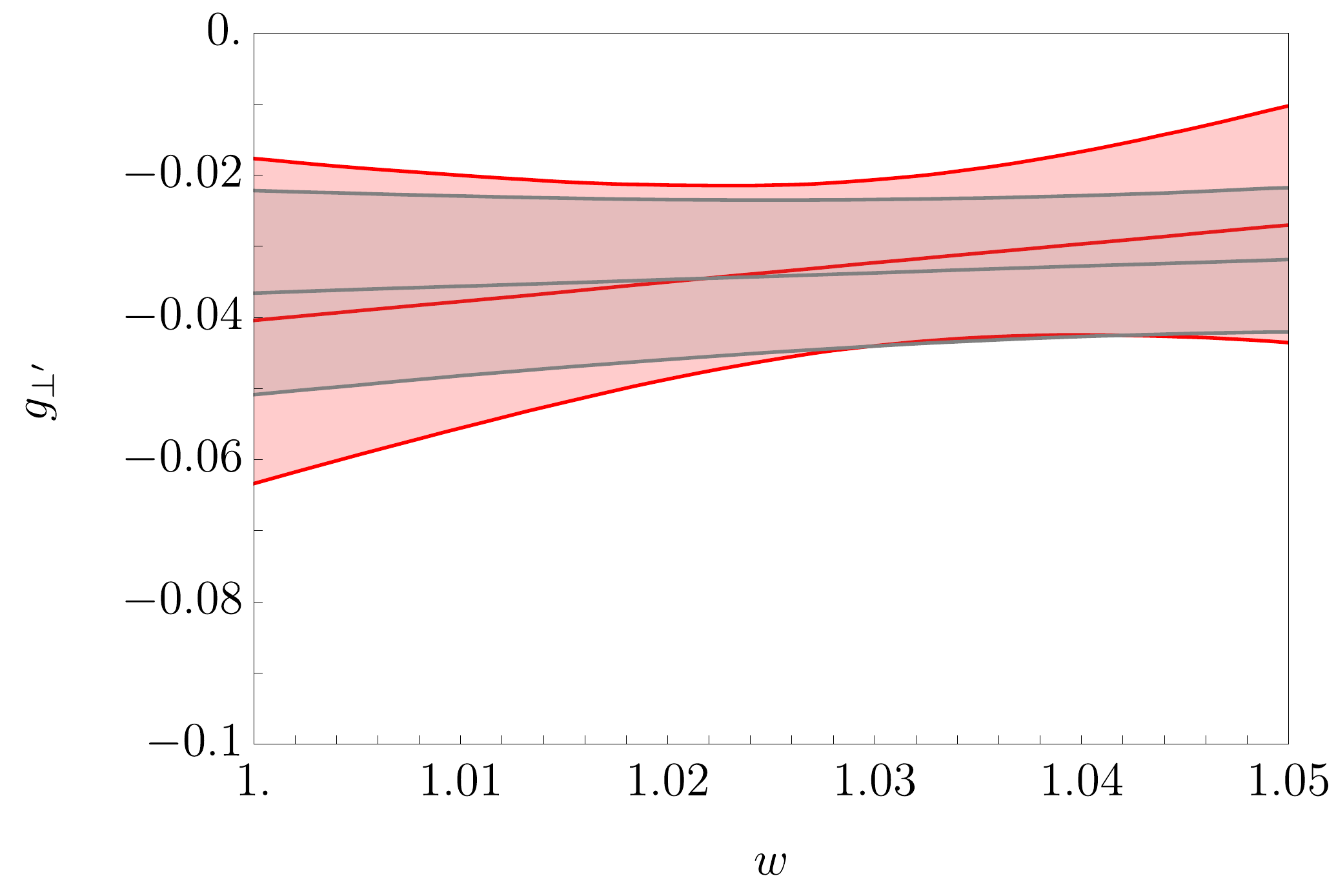}  \\[1.25em]
       \includegraphics[scale=0.32]{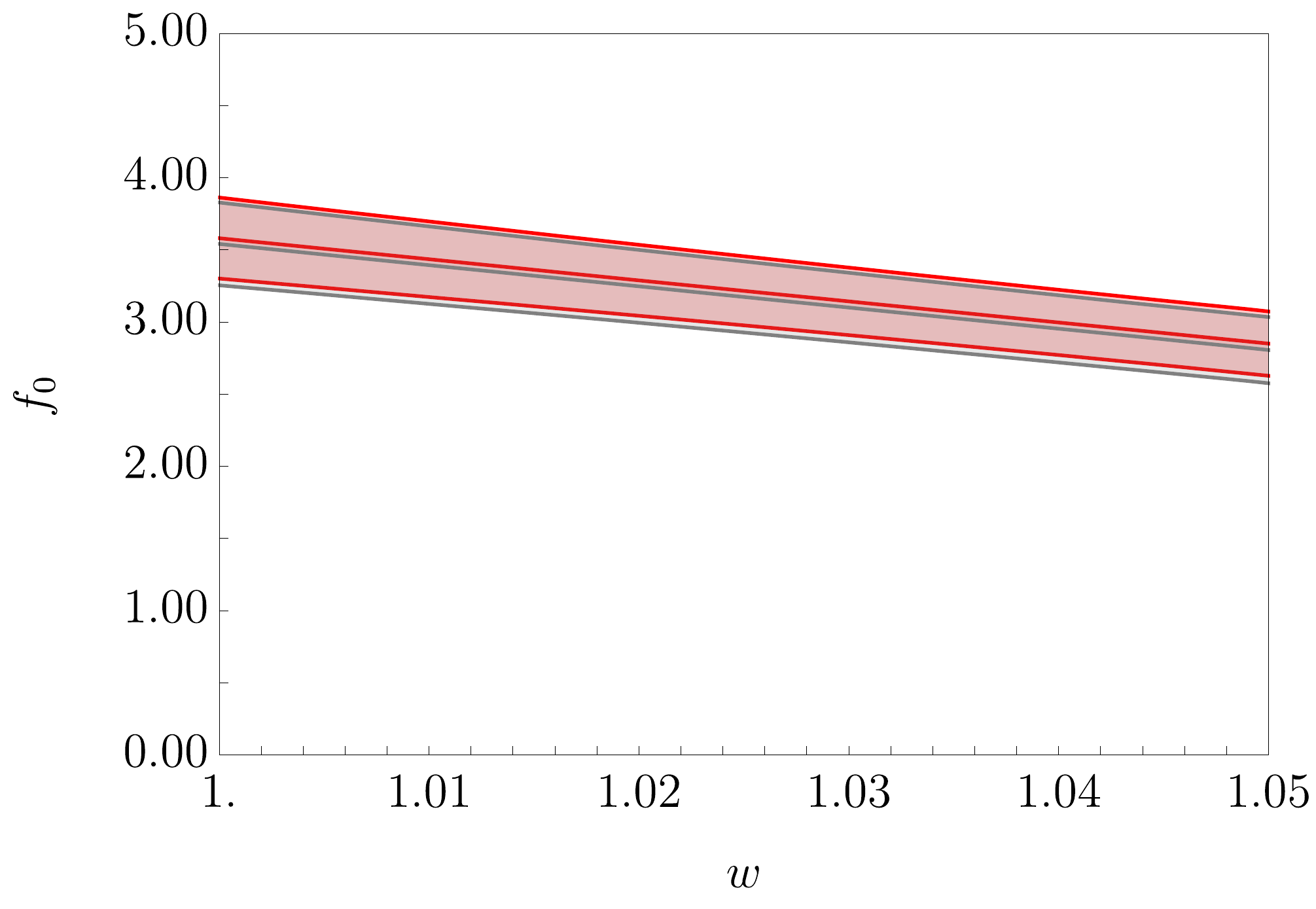}  &
       \includegraphics[scale=0.32]{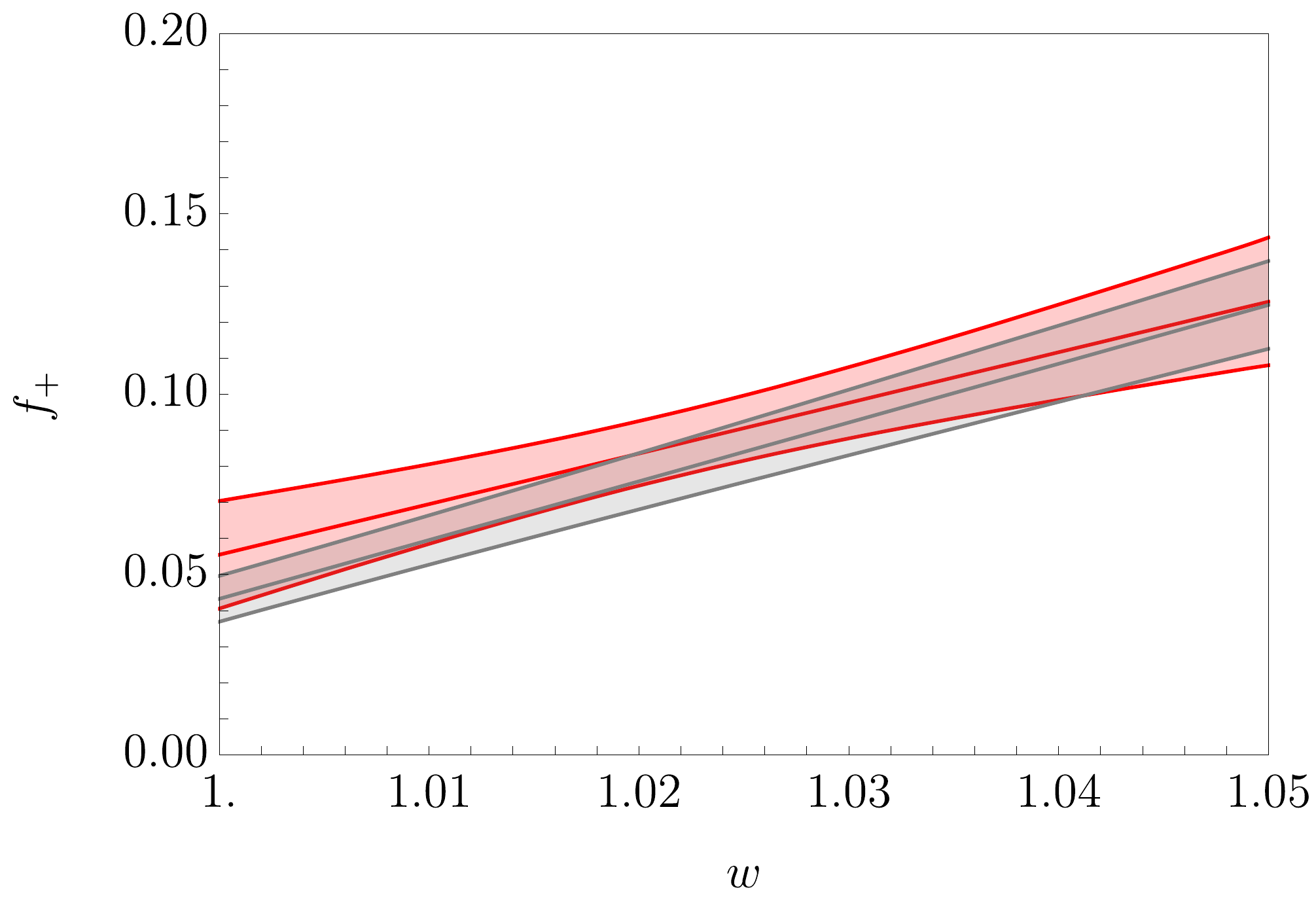}  \\[1.25em]
        \includegraphics[scale=0.32]{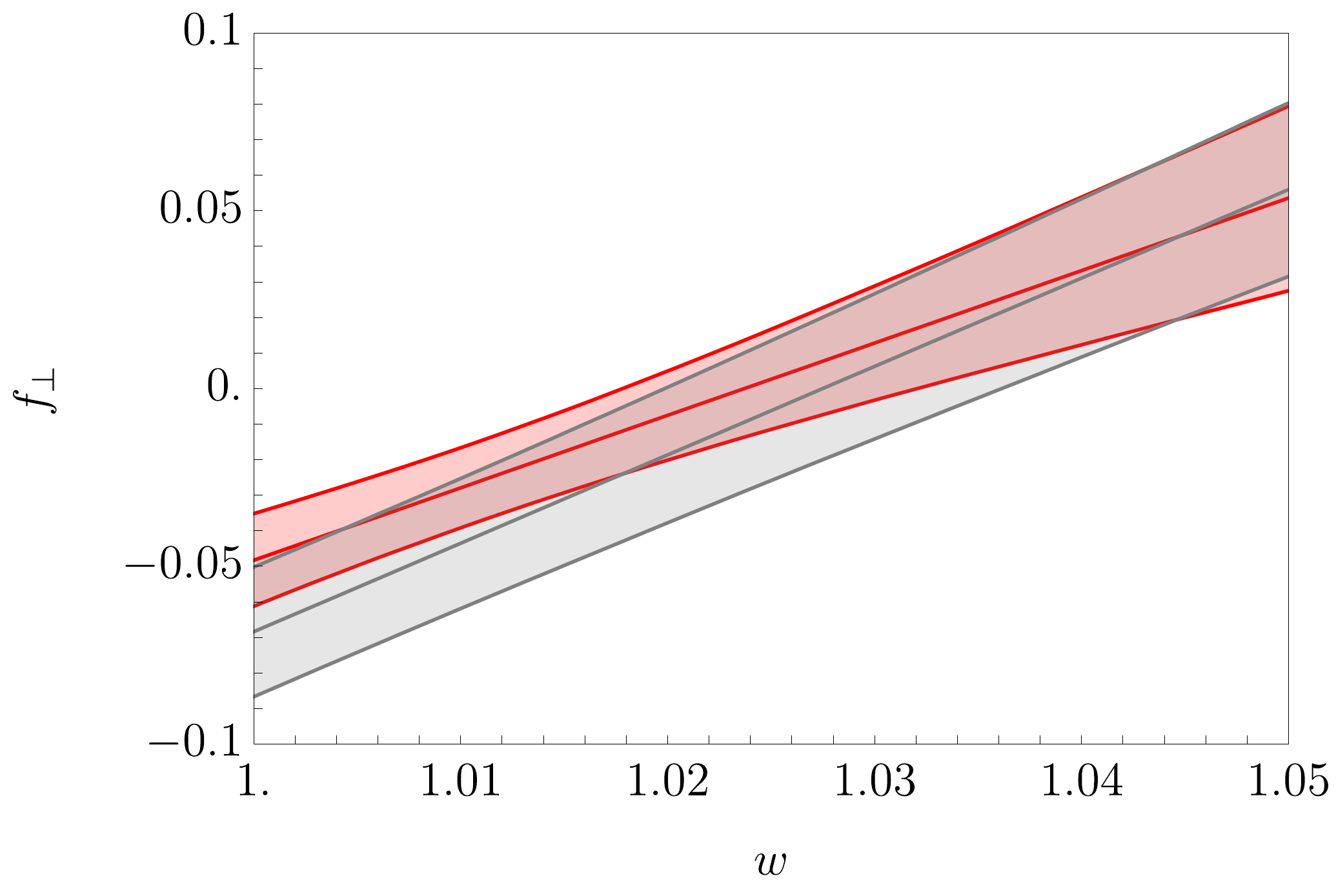}   &
       \includegraphics[scale=0.32]{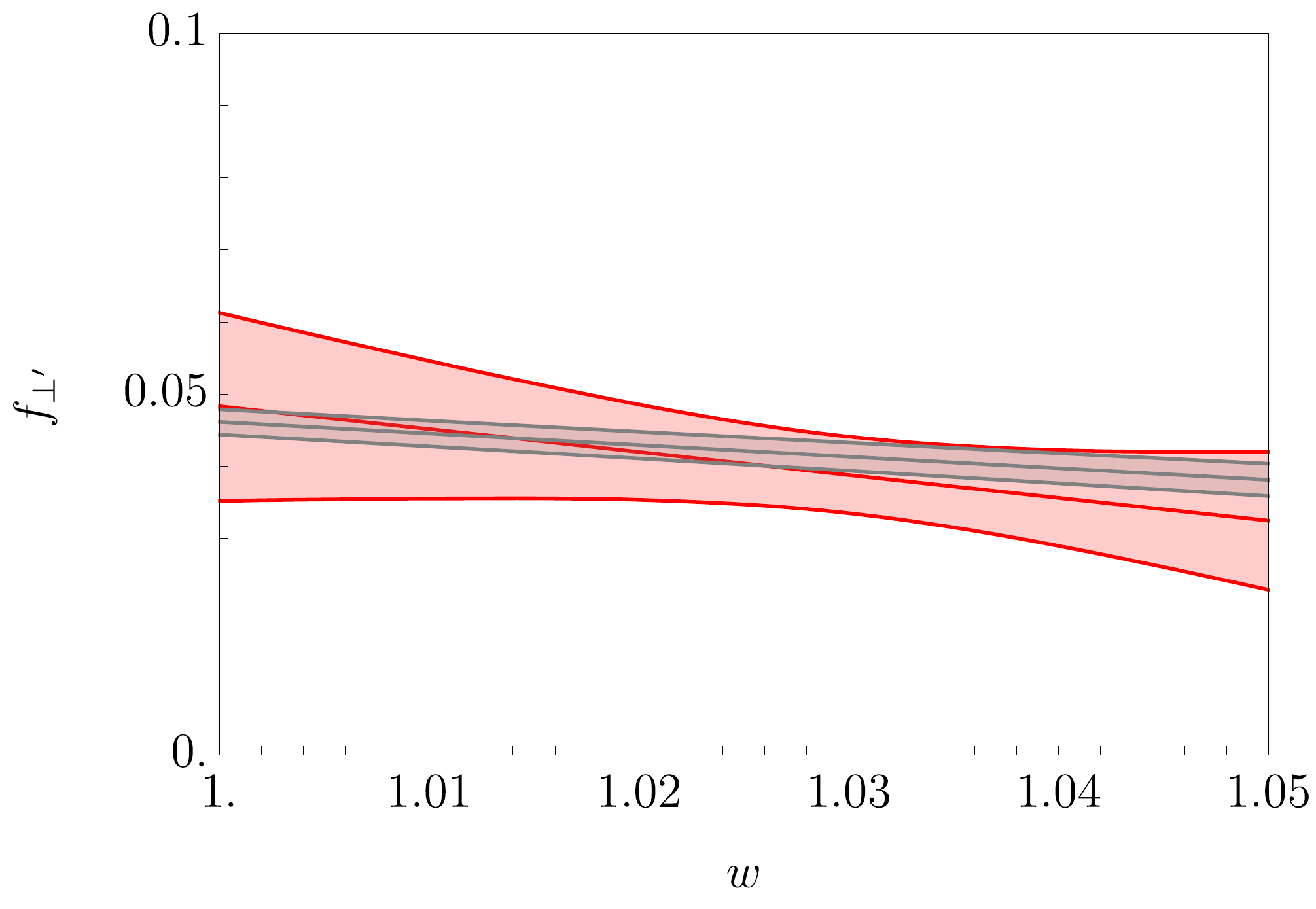}  
    \end{tabular}
    \end{center}
\caption{Comparison between the lattice results in Ref.~\cite{Meinel:2020owd} (grey band) and the fit results for the HQE form factors (red band) for the vector and axial-vector form factors. The two bands represent the $68\%$ interval.}
\label{fig:ffs}
\end{figure}

The best-fit point for the hadronic parameters and their uncertainties is shown in \Table{tab:fit_values}, and the correlation matrix among the parameters is given in \app{app:C}. With this result, I compare the HQE form factors to the lattice QCD ones. The comparison is given in \fig{fig:ffs}, showing excellent agreement between the two parametrisations. \\

\begin{figure}[H]
\begin{center}
  \begin{tabular}{cc}
       \includegraphics[scale=0.32]{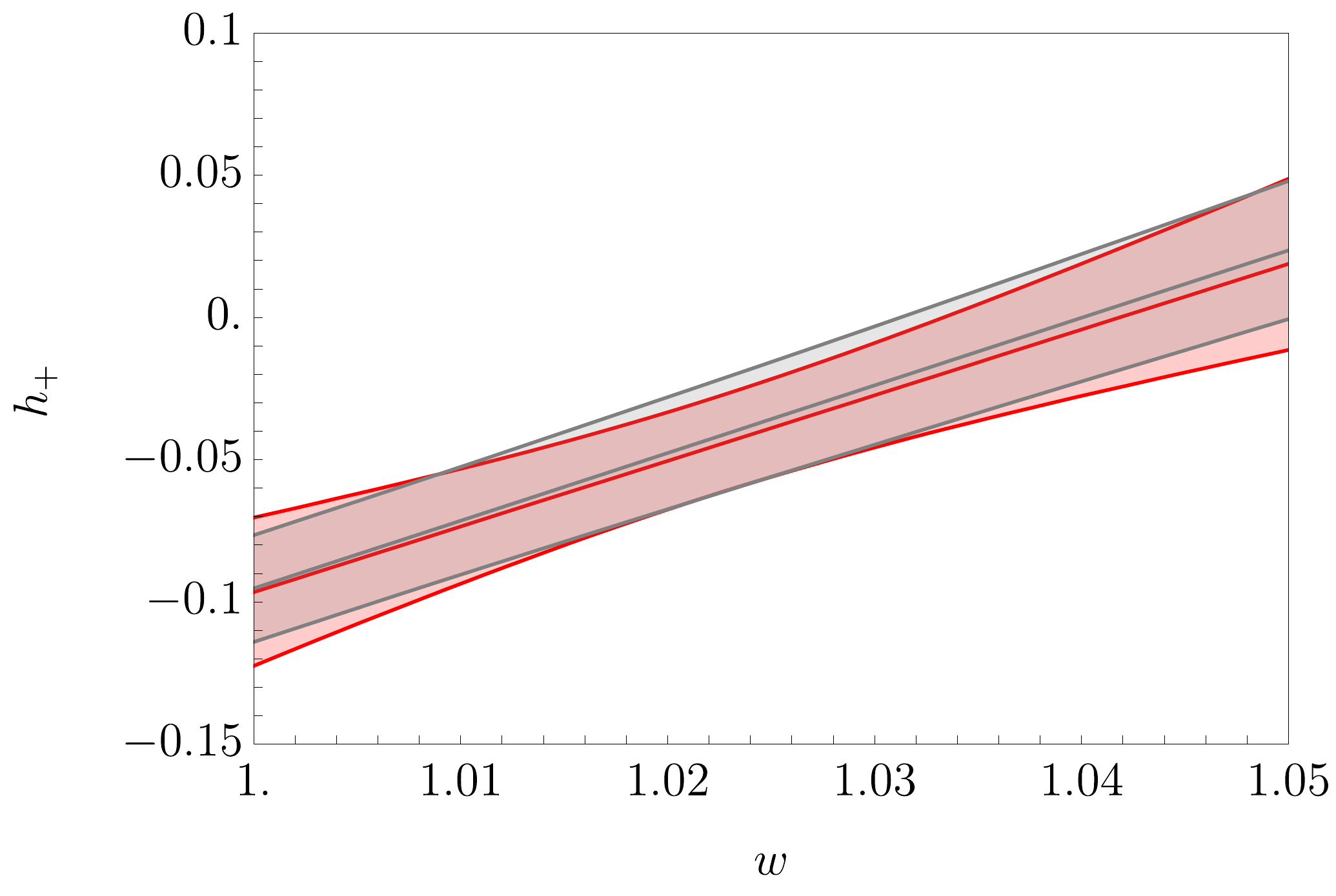}  &
        \includegraphics[scale=0.32]{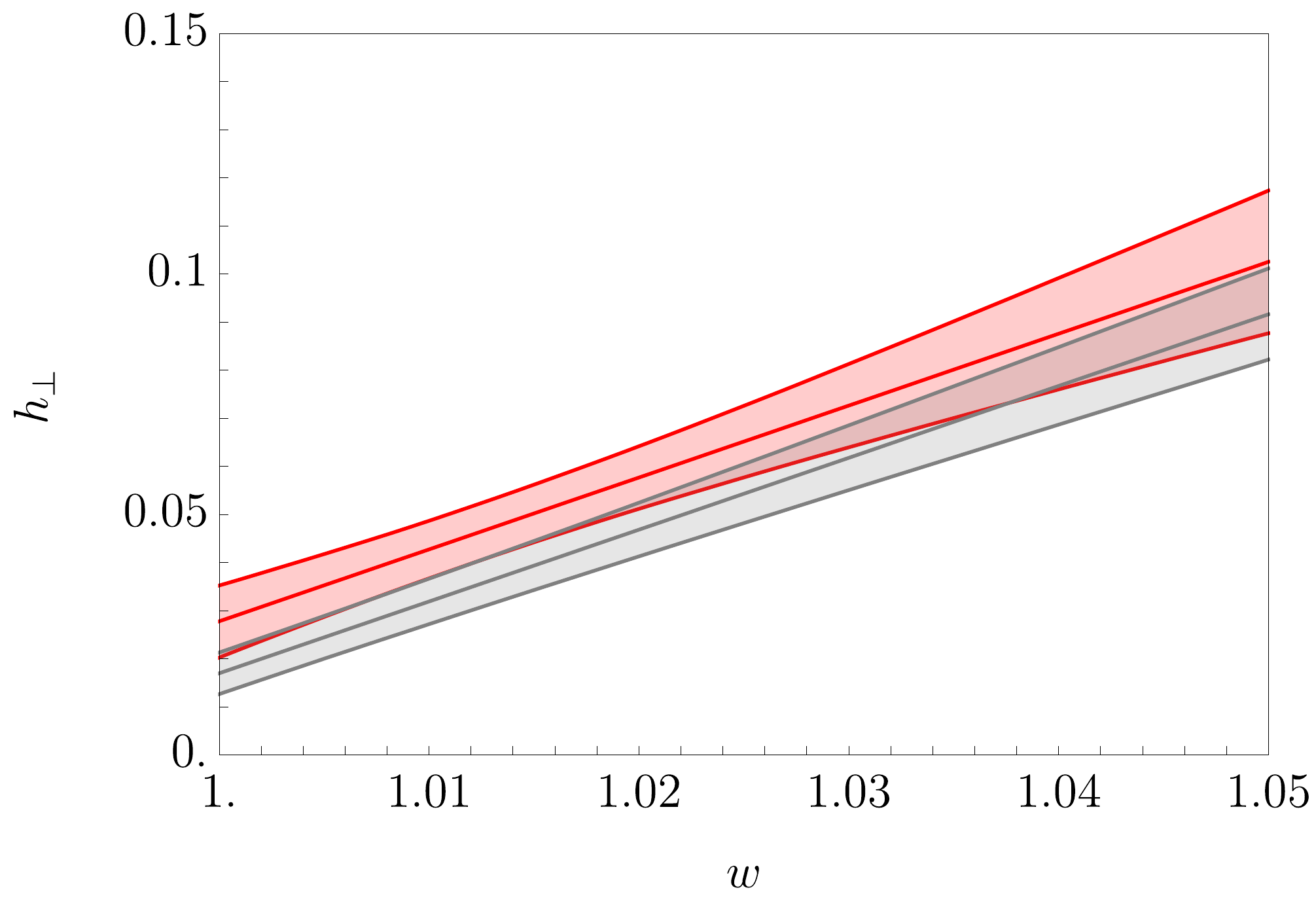} \\[1.25em]
        \includegraphics[scale=0.32]{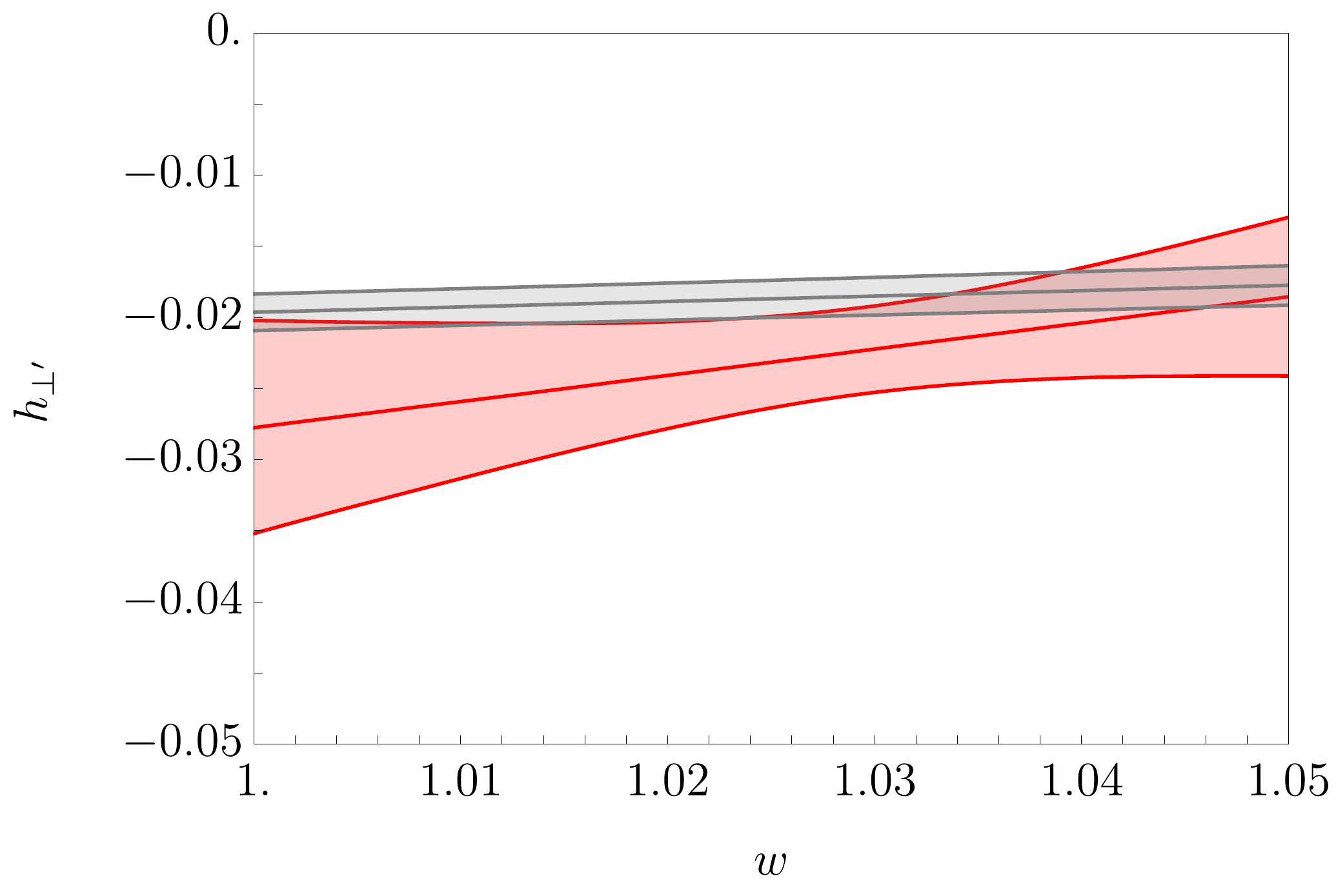}  &
       \includegraphics[scale=0.32]{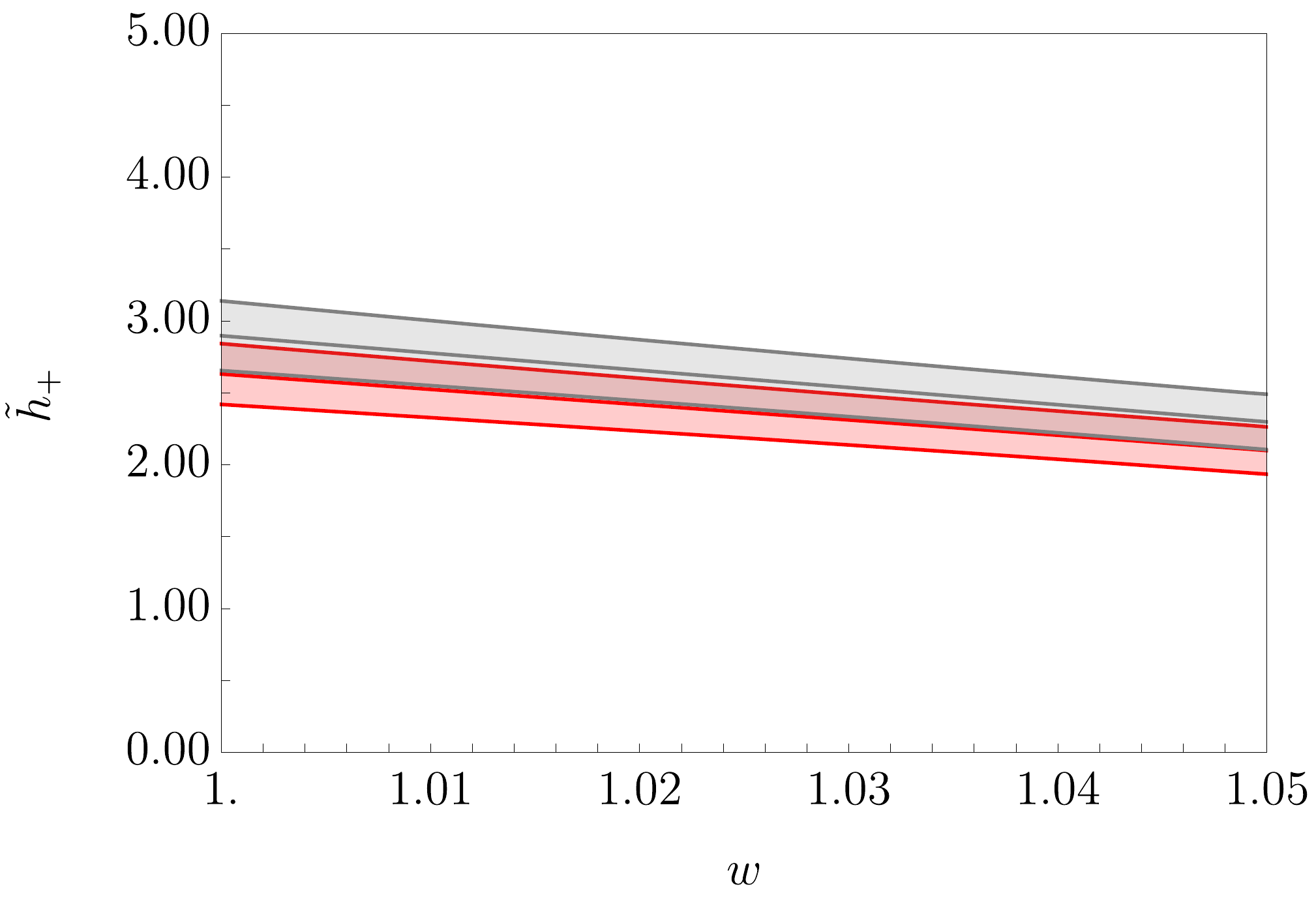}  \\[1.25em]
        \includegraphics[scale=0.32]{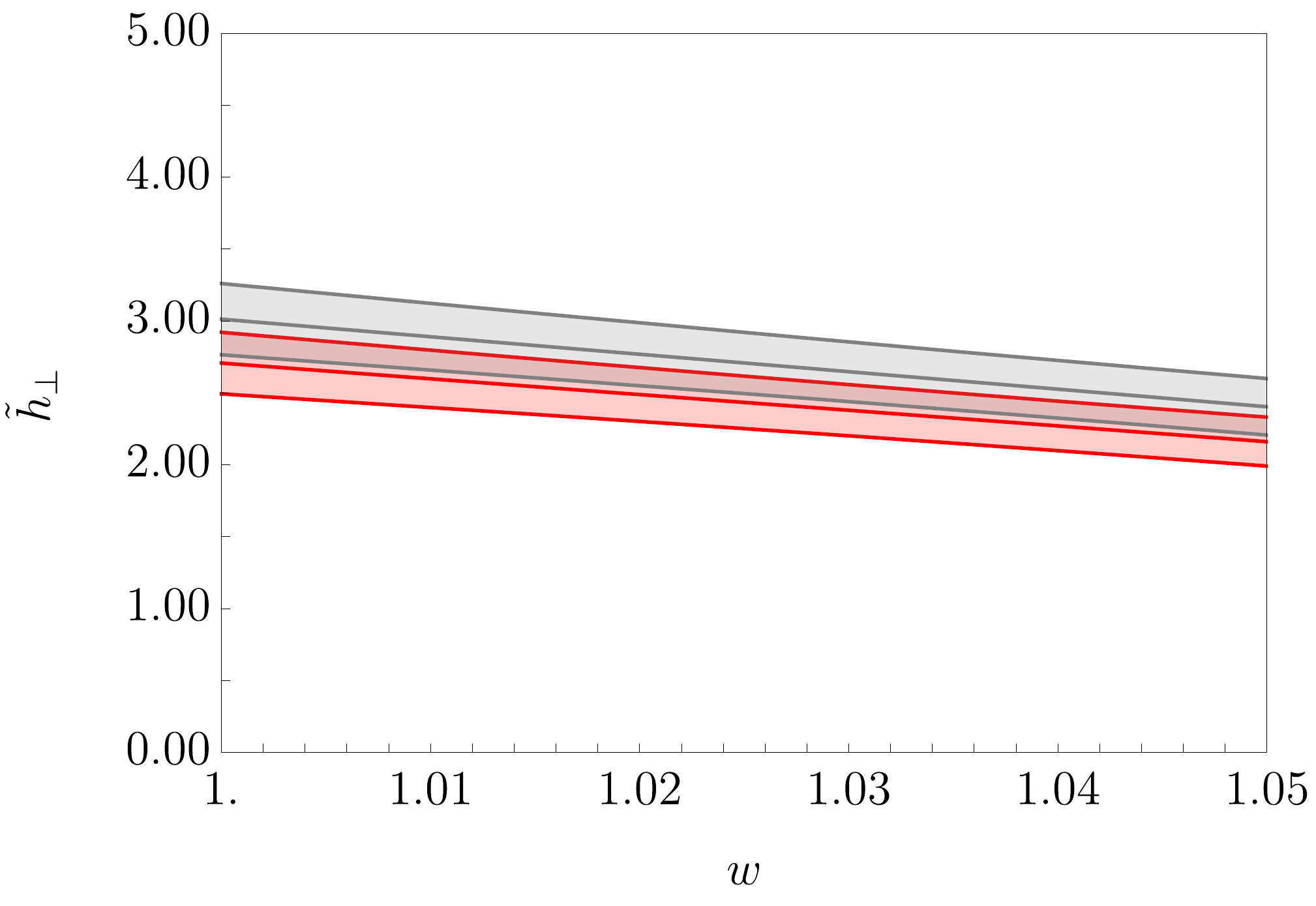}   &
       \includegraphics[scale=0.32]{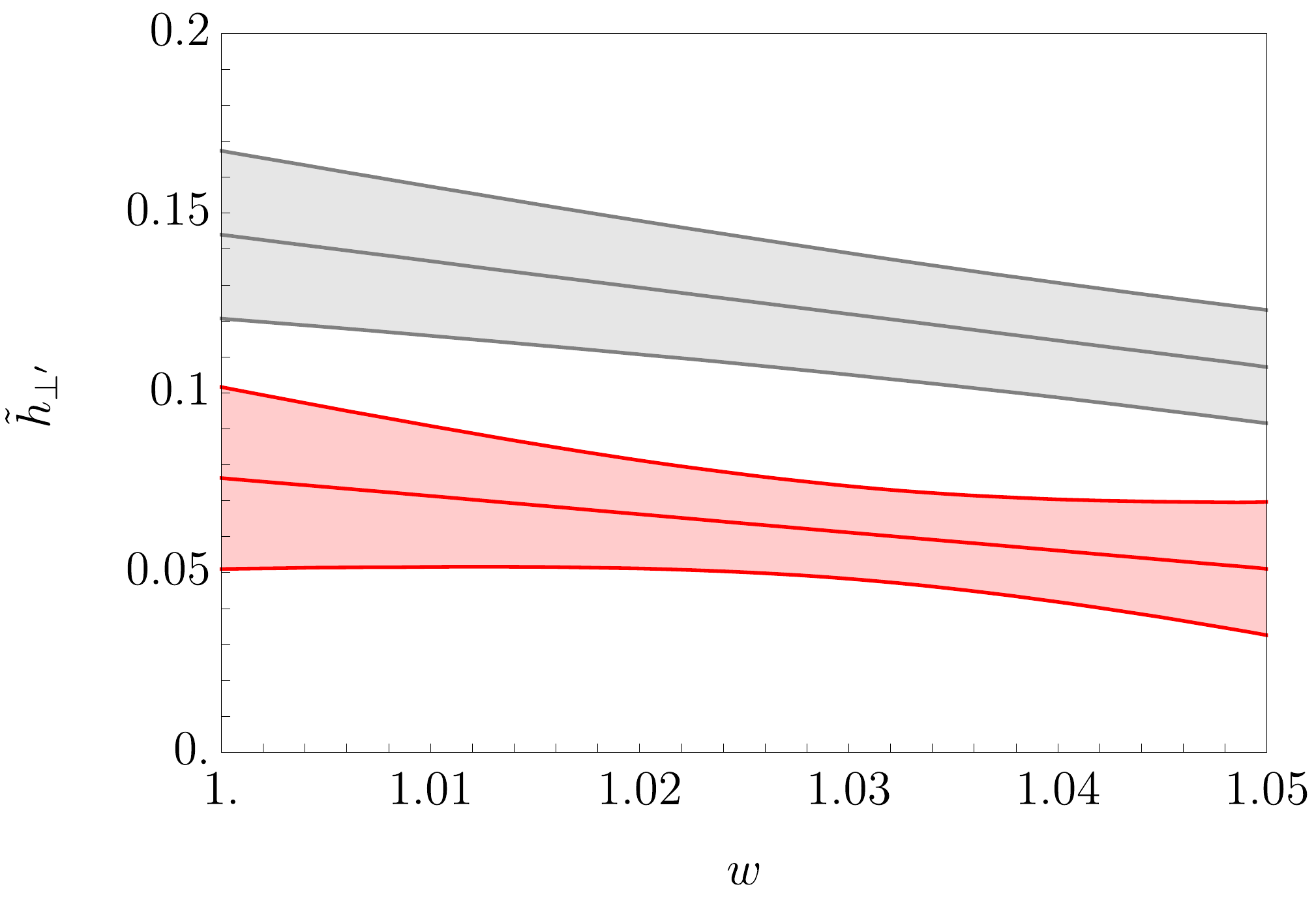}  
    \end{tabular}
    \end{center}
\caption{Comparison between the lattice results in Ref.~\cite{Meinel:2020owd} (grey band) and the predictions for tensor and pseudo-tensor form factors based on the fir results in \Table{tab:fit_values} (red band). The two bands represent the $68\%$ interval.}
\label{fig:ffstens}
\end{figure}

I use the results of the fit to obtain predictions for the tensor and pseudo-tensor form factors. The comparison between them and the lattice QCD computation is shown in \fig{fig:ffstens} \footnote{The lattice QCD results employed here for tensor and pseudo-tensor form factors differ by a global sign with respect to the first version of Ref.\cite{Meinel:2020owd}. This sign inconsistency will be fixed in a forthcoming update of Ref.\cite{Meinel:2020owd}.}. 
The form factors $h_{\perp^\prime}$ and $\tilde h_{\perp^\prime}$ show a tension between lattice QCD data and HQE predictions. This tension manifests itself also when introducing the tensor and pseudo-tensor form factors in the fit. This procedure increases the $\chi^2/\text{d.o.f.}$, making it much higher than 1. This corroborates the choice of excluding them from the fit procedure. \\

\begin{figure}
\centering
\subfloat{\includegraphics[scale = 0.32]{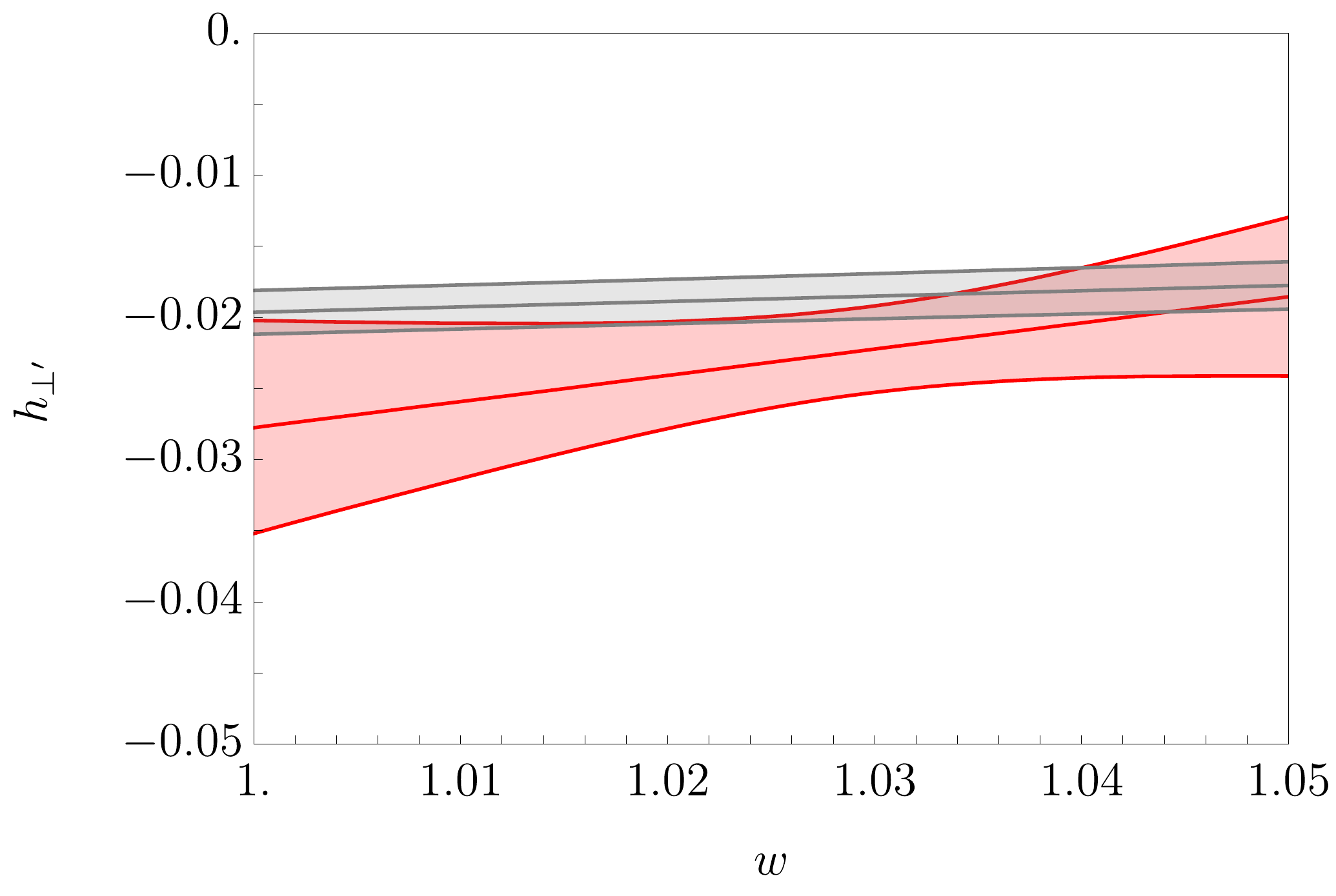}}
\hspace{5mm}
\subfloat{\includegraphics[scale = 0.32]{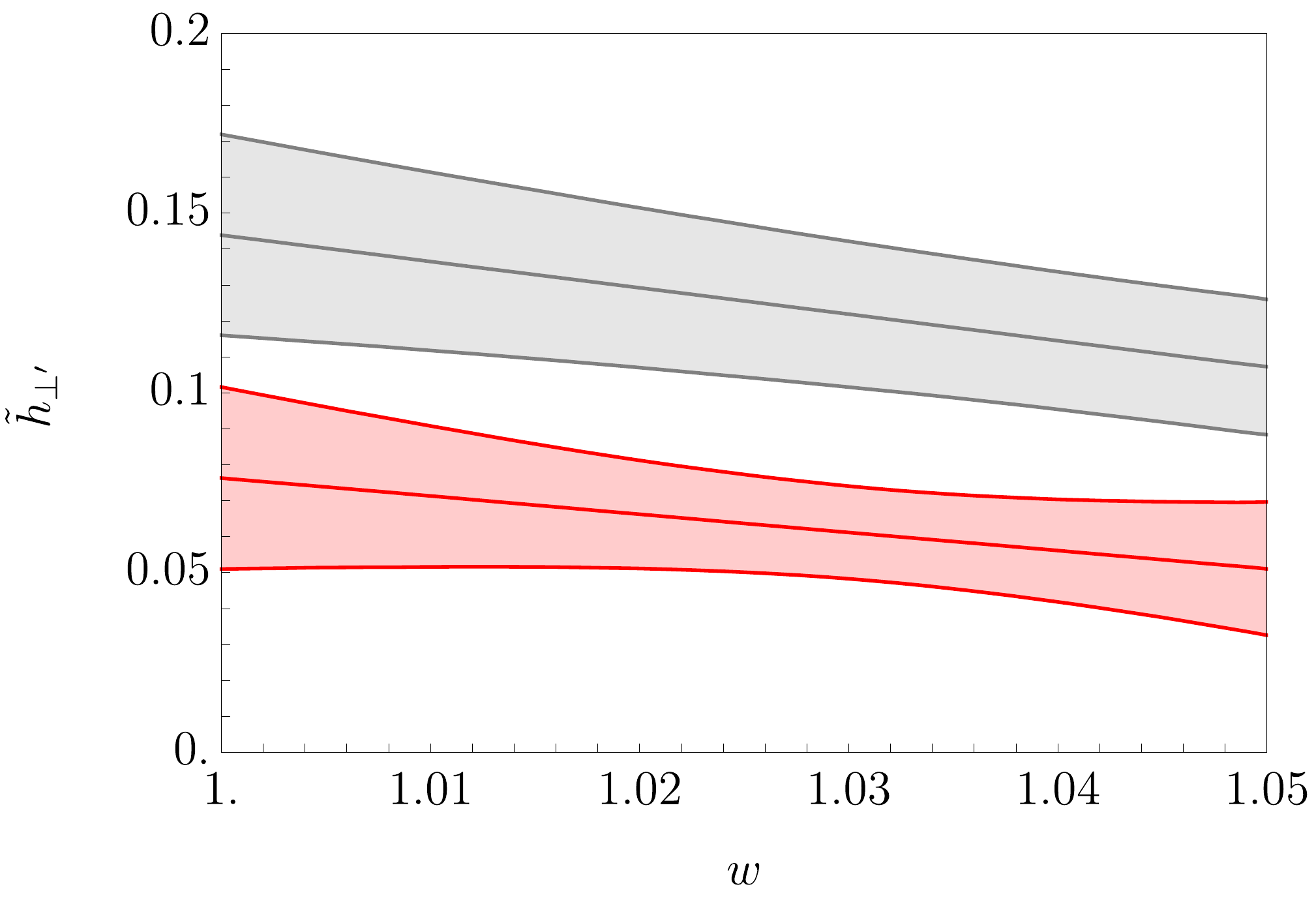}}
\caption{Comparison between the lattice QCD results in Ref.~\cite{Meinel:2020owd} (grey band) and the predictions for tensor and pseudo-tensor form factors based on the fir results in \Table{tab:fit_values} (red band) with lattice QCD uncertainties inflated of $20\%$. The two bands represent the $68\%$ interval.}
\label{fig:ffstens_infl}
\end{figure}

Sources of these tensions can be looked for in the lattice QCD data and  HQE framework. The hypotheses are mainly two: \textit{i)} the uncertainties on the lattice QCD parameters describing tensor and pseudo-tensor form factors are underestimated, and \textit{ii)} missing corrections in the HQE cause a shift in the hadronic parameters. In the case of \textit{i)}, I checked explicitly the result of inflating lattice QCD uncertainties by $20\%$ for $h_{\perp^\prime}$ and $\tilde h_{\perp^\prime}$. In \fig{fig:ffstens_infl} the results of this test are shown, proving that the compatibility slightly improves, even if it is still poor in the case of $\tilde h_{\perp^\prime}$. Concerning \textit{ii)}, the most important corrections in the HQE, beside the ones already discussed in this work, arise at order $\mathcal{O}(\alpha_s/m_b)$, and could produce a $\mathcal{O}(\text{few}\,\%)$ shift in the central value of the form factors parameters. However, assessing the quantitative impact of these corrections requires understanding how they affect the form factors in a correlated way. \\
From these estimates, it seems that neither \textit{i)} nor \textit{ii)} can explain the tension on their own.  Most likely, a combination of the two effects might help to reconcile the HQE for the tensor and pseudo-tensor form factors with the current lattice QCD data.

\section{Conclusions}
\label{sec:4}
I revisit the Heavy Quark Expansion of the $\Lb\to\Lst(1520)$ form factors including next-to-leading order $\alpha_s$ corrections and for the first time next-to-leading power $1/m_b$ corrections. In this framework, form factors depend on unknown hadronic parameters which are obtained by fitting the form factor parametrisation here discussed to a recent lattice QCD computation \cite{Meinel:2020owd}. I perform the fit using data for vector and axial-vector form factors, finding good agreement between the lattice QCD calculation and the Heavy Quark Expansion predictions. The fit results are used to predict tensor and pseudo-tensor form factors, showing tensions between the Heavy Quark Expansion-based predictions and the lattice QCD data.
I discuss two possible sources of the tensions: an underestimation of the uncertainties on the lattice QCD parameters involved in these form factors and missing higher-order terms in the Heavy Quark Expansion, e.g. at order $\mathcal{O}(\alpha_s/m_b)$. Most likely, only a combination of these two effects could reconcile lattice QCD determination and Heavy Quark Expansion based parametrisation of tensor and pseudo-tensor form factors. Until then, it is not possible to claim high precision in the Heavy Quark Expansion parametrisation of $\Lb\to\Lst(1520)$ form factors.\\
Besides this, I want to point out the need to extend the calculation of the $\Lb\to\Lst(1520)$ form factors to the large-recoil region. Quark models \cite{Mott:2011cx} are available, although without a consistent treatment of uncertainties. It is therefore needed to perform up-to-date calculations of the $\Lb\to\Lst(1520)$ form factors using e.g. sum rules at $q^2 \leq 0$. Estimates of this type will allow to extrapolate the form factors to the large-recoil region and to assess the magnitude of their curvature. These studies will be crucial for future LHCb analysis of $\Lb\to\Lst(1520)\ell^+\ell^-$ decays.


\vspace{6pt}


\section*{Acknowledgments}I thank Stefan Meinel and Gumaro Rendon for insightful discussions on lattice QCD data. I am also grateful to Bernat Capdevila, Thorsten Feldmann, Paolo Gambino, Martin Jung and Danny van Dyk for useful discussions. I also acknowledge useful inputs on the text from Nico Gubernari.\\
This work is supported by Deutsche Forschungsgemeinschaft (DFG, German Research Foundation) under grant 396021762 - TRR 257 ``Particle Physics Phenomenology after the Higgs Discovery'' and by the Italian Ministry of Research (MIUR) under grant PRIN 20172LNEEZ. \\

\appendix
\section{Details on the form factor parametrisation}
\label{app:A}
Concerning the vector and the axial vector currents, we follow the notation of Ref.\cite{Boer:2018vpx}. For the vector current I have:
\begin{align}
    \Gamma_{V,(1/2,t)}^{\alpha\mu} & = \frac{\sqrt{4 \mLb \mLst}}{\sqrt{s_+}}\, \frac{2 \mLst}{\sqrt{s_ +s_-}} p^\alpha\,\frac{\mLb - \mLst}{\sqrt{q^2}} \, \frac{q^\mu}{\sqrt{q^2}}\,,\\
\Gamma_{V,(1/2,0)}^{\alpha\mu}     & = \frac{\sqrt{4 \mLb \mLst}}{\sqrt{s_-}}\, \frac{2 \mLst}{\sqrt{s_ +s_-}} p^\alpha\,\frac{\mLb + \mLst}{s_+}\left[(p + k)^\mu - \frac{\mLb^2 - \mLst^2}{q^2} q^\mu\right]\,,\\
\Gamma_{V,(1/2,\perp)}^{\alpha\mu} & = \frac{\sqrt{4 \mLb \mLst}}{\sqrt{s_-}}\, \frac{2 \mLst}{\sqrt{s_ +s_-}} p^\alpha\,\left[\gamma^\mu - \frac{2 \mLst}{s_+} p^\mu - \frac{2 \mLb}{s_+} k^\mu\right]\,,\\
\Gamma_{V,(3/2,\perp)}^{\alpha\mu} & = \frac{\sqrt{4 \mLb \mLst}}{\sqrt{s_-}}\, \frac{-4 i \epsilon^{\alpha\mu p k}}{\sqrt{s_+ s_-}} \gamma_5 + \Gamma_{V,(1/2,\perp)}\,,
\end{align}
while for the axial vector current:
\begin{align}
\Gamma_{A,(1/2,t)}^{\alpha\mu} & = \frac{\sqrt{4 \mLb \mLst}}{\sqrt{s_-}}\, \frac{2 \mLst}{\sqrt{s_ +s_-}} p^\alpha\,\frac{\mLb + \mLst}{\sqrt{q^2}} \, \frac{q^\mu}{\sqrt{q^2}}\,,\\
\Gamma_{A,(1/2,0)}^{\alpha\mu}     & = \frac{\sqrt{4 \mLb \mLst}}{\sqrt{s_+}}\, \frac{2 \mLst}{\sqrt{s_ +s_-}} p^\alpha\,\frac{\mLb - \mLst}{s_-}\left[(p + k)^\mu - \frac{\mLb^2 - \mLst^2}{q^2} q^\mu\right]\,,\\
\Gamma_{A,(1/2,\perp)}^{\alpha\mu} & = \frac{\sqrt{4 \mLb \mLst}}{\sqrt{s_+}}\, \frac{2 \mLst}{\sqrt{s_ +s_-}} p^\alpha\,\left[\gamma^\mu + \frac{2 \mLst}{s_-} p^\mu - \frac{2 \mLb}{s_-} k^\mu\right]\,,\\
\Gamma_{A,(3/2,\perp)}^{\alpha\mu} & = \frac{\sqrt{4 \mLb \mLst}}{\sqrt{s_+}}\, \frac{-4 i \epsilon^{\alpha\mu p k}}{\sqrt{s_+ s_-}} \gamma_5 - \Gamma_{A,(1/2,\perp)}\,.
\end{align}
Concerning the tensor currents, we modify the parametrisation in Ref.~\cite{Descotes-Genon:2019dbw} by rescaling each structure with suitable factors. We have:
\begin{align}
\Gamma_{T,(1/2,0)}^{\alpha\mu}     & = \frac{\sqrt{4 \mLb \mLst}}{\sqrt{s_+}}\, \frac{q^2}{s_+ s_-} p^\alpha\left[(p + k)^\mu - \frac{\mLb^2 - \mLst^2}{q^2} q^\mu\right]\,,\\
\Gamma_{T,(1/2,\perp)}^{\alpha\mu} & = \frac{\sqrt{4 \mLb \mLst}}{\sqrt{s_+}}\, \frac{\mLb+\mLst}{s_-} p^\alpha\,\left[\gamma^\mu -2\frac{\mLst}{s_+} p^\mu  -2\frac{\mLb}{s_+} k^\mu\right]\,,\\
\Gamma_{T,(3/2,\perp)}^{\alpha\mu} & = \frac{\sqrt{4 \mLb \mLst}}{\sqrt{s_+}}\, \left[g^{\alpha\mu}+\frac{\mLst}{s_-}p^\alpha\left(\gamma^\mu-2\frac{1}{\mLst} k^\mu+ 2\frac{\mLst}{s_+}p^\mu +2\frac{\mLb}{s_+}k^\mu\right)\right]\,,
\label{eq:structT}
\end{align}
and 
\begin{align}
\Gamma_{T5,(1/2,0)}^{\alpha\mu}     & = \frac{\sqrt{4 \mLb \mLst}}{\sqrt{s_-}}\, \frac{q^2}{s_+ s_-} p^\alpha\left[(p + k)^\mu - \frac{\mLb^2 - \mLst^2}{q^2} q^\mu\right]\,,\\
\Gamma_{T5,(1/2,\perp)}^{\alpha\mu} & = \frac{\sqrt{4 \mLb \mLst}}{\sqrt{s_-}}\, \frac{\mLb-\mLst}{s_+} p^\alpha\,\left[\gamma^\mu +2\frac{\mLst}{s_-} p^\mu  -2\frac{\mLb}{s_-} k^\mu\right]\,,\\
\Gamma_{T5,(3/2,\perp)}^{\alpha\mu} & = \frac{\sqrt{4 \mLb \mLst}}{\sqrt{s_-}}\, \left[g^{\alpha\mu}-\frac{\mLst}{s_+}p^\alpha\left(\gamma^\mu+2\frac{1}{\mLst} k^\mu-2\frac{\mLst}{s_-}p^\mu +2\frac{\mLb}{s_-}k^\mu\right)\right]\,.
\label{eq:structT5}
\end{align}
I define the helicity amplitudes as 
\begin{equation}
\mathcal{A}_\Gamma(s_b,s_\Lambda,\lambda_\Lambda,\lambda_q)= \bra{\Lst (s_\Lambda,\eta(\lambda_\Lambda))}\bar{s}\Gamma^\mu\epsilon^*_\mu(\lambda_q)b\ket{ \Lb(s_b) }\,,
\end{equation}
where $\epsilon^*_\mu(\lambda_q)$ are a basis of polarisation vectors for the virtual W exchange with polarisation states $\lambda_q = \{t,0,+1,-1\}$. For details see Ref.~\cite{Boer:2018vpx}.
The physical helicity amplitudes are identified by the following set:
\begin{equation}
\label{eq:def:hel-amp32}
\begin{aligned}
    \mathcal{A}_\Gamma(+1/2, +3/2, +1) & \equiv \mathcal{A}_\Gamma(+1/2, +1/2, +1, +1)\,,\\
    \mathcal{A}_\Gamma(+1/2, +1/2,  0) & \equiv \sqrt{\frac{2}{3}} \mathcal{A}_\Gamma(+1/2, +1/2, 0, 0) + \sqrt{\frac{1}{3}} \mathcal{A}^{(3/2)}_\Gamma(+1/2, -1/2, +1, 0)\,,\\
    \mathcal{A}_\Gamma(+1/2, +1/2,  t) & \equiv \sqrt{\frac{2}{3}} \mathcal{A}_\Gamma(+1/2, +1/2, 0, t) + \sqrt{\frac{1}{3}} \mathcal{A}^{(3/2)}_\Gamma(+1/2, -1/2, +1, t)\,,\\
    \mathcal{A}_\Gamma(+1/2, -1/2, -1) & \equiv \sqrt{\frac{2}{3}} \mathcal{A}_\Gamma(+1/2, -1/2, 0,-1) + \sqrt{\frac{1}{3}} \mathcal{A}^{(3/2)}_\Gamma(+1/2, +1/2, -1,-1)\,,
\end{aligned}
\end{equation}
where $\Gamma$ represents the four possible currents. In the case of the vector and axial vector current, the helicity amplitudes are already calculated in Ref.~\cite{Boer:2018vpx}. For convenience, I adapt them to this case and report them here:
\begin{align}
\mathcal{A}_V(+1/2, +3/2, +1)&= - 4 \sqrt{\mLb \mLst} F_{3/2,\perp} \,, \label{eq:HAVQCD:1}\\
\mathcal{A}_V(+1/2, +1/2,  0) &= 2 \sqrt{\frac{2 \, \mLb \mLst}{3\, q^2}} (\mLb+\mLst)F_{1/2,0}\,, \label{eq:HAVQCD:2}\\
\mathcal{A}_V(+1/2, +1/2,  t) &= 2 \sqrt{\frac{2 \, \mLb \mLst}{3\, q^2}} (\mLb-\mLst)F_{1/2,t}\,, \label{eq:HAVQCD:3}\\
\mathcal{A}_V(+1/2, -1/2, -1) &=  -\frac{4}{\sqrt{3}}\sqrt{\mLb\mLst}F_{1/2,\perp}\,,\label{eq:HAVQCD:4}
\end{align}
and for the axial vector current:
\begin{align}
\mathcal{A}_A(+1/2, +3/2, +1)&= - 4 \sqrt{\mLb \mLst} G_{3/2,\perp}\,, \label{eq:HAAQCD:1}\\
\mathcal{A}_A(+1/2, +1/2,  0) &= 2 \sqrt{\frac{2 \, \mLb \mLst}{3\, q^2}} (\mLb-\mLst)G_{1/2,0}\,, \label{eq:HAAQCD:2}\\
\mathcal{A}_A(+1/2, +1/2,  t) &= 2 \sqrt{\frac{2 \, \mLb \mLst}{3\, q^2}} (\mLb+\mLst)G_{1/2,t}\,, \label{eq:HAAQCD:3}\\
\mathcal{A}_A(+1/2, -1/2, -1) &=  \frac{4}{\sqrt{3}}\sqrt{\mLb\mLst}G_{1/2,\perp}\,. \label{eq:HAAQCD:4}
\end{align}
In the case of tensor currents, with the definitions in \eqs{eq:structT}{eq:structT5}, I find
\begin{align}
    \mathcal{A}_T(+1/2, +3/2, +1) &= -2 \sqrt{\mLb \mLst} \,T_{3/2,\perp}\,, \label{eq:HATQCD:1}\\
    \mathcal{A}_T(+1/2, +1/2,  0) & =-\sqrt{\frac{2}{3}}\sqrt{\frac{\mLb}{\mLst}q^2} \,T_{1/2,0}\,,\label{eq:HATQCD:2}\\
    \mathcal{A}_T(+1/2, -1/2, -1) & =+\frac{2}{\sqrt{3}}\sqrt{\frac{\mLb}{ \mLst}}(\mLb+\mLst) T_{1/2,\perp}\,, \label{eq:HATQCD:3}\\
    \mathcal{A}_T(+1/2, +1/2,  t) &= 0\,,
\end{align}
and
\begin{align}
    \mathcal{A}_{T5}(+1/2, +3/2, +1) &= +2 \sqrt{\mLb \mLst}\, T^5_{3/2,\perp}\,,\label{eq:HAT5QCD:1}\\
    \mathcal{A}_{T5}(+1/2, +1/2,  0) & =+\sqrt{\frac{2}{3}}\sqrt{\frac{\mLb}{\mLst}q^2} \,T^5_{1/2,0}\,, \label{eq:HAT5QCD:2}\\
    \mathcal{A}_{T5}(+1/2, -1/2, -1) & =+\frac{2}{\sqrt{3}}\sqrt{\frac{\mLb}{ \mLst}}(\mLb-\mLst) T_{1/2,\perp}\,, \label{eq:HAT5QCD:3}\\
    \mathcal{A}_{T5}(+1/2, +1/2,  t) &=0\,.
\end{align}
In the heavy quark expansion, the helicity amplitudes concerning the vector current read:
\begin{align}
\mathcal{A}_V(+1/2, +3/2, +1)&= 2 \frac{s_+}{\mLb} (\zeta_1^\text{SL}+\zeta_2^\text{SL}) \label{eq:HAVHQE:1} \,, \\
\mathcal{A}_V(+1/2, +1/2,  0) &= \frac{\sqrt{2s_+}}{\mLb\mLst\sqrt{3q^2}} \bigg\{\frac{\sMinus}{\mLb} \left[(1+ C_0^{(v)}) (\mLst^2+\mLst\mLb-q^2)+\frac{1}{2}C_1^{(v)} \sPlus\right]\zeta_2 \nonumber \\
+&\sMinus\left[(1+ C_0^{(v)}) (\mLst+\mLb)+\frac{1}{2\mLb}C_1^{(v)} \sPlus\right]\zeta_1 \nonumber\\
-&\left[\mLb^2(\mLst^2-\mLb^2+q^2)+\sMinus\sPlus\right]\zeta_1^\text{SL}-(\mLst^2-\mLb^2+q^2)\zeta_2^\text{SL}\nonumber\\
-&\frac{\sMinus}{2\mLb^2}\left[(2\mLst^2+3 \mLst\mLb+\mLb^2-2 q^2)\zeta_3^\text{SL}-(\mLst^2+3 \mLst\mLb+2\mLb^2-q^2)\zeta_4^\text{SL}\right]\bigg\}\,,\label{eq:HAVHQE:2}  \\
\mathcal{A}_V(+1/2, +1/2,  t) &= \frac{\sqrt{2\sMinus}\sPlus}{\mLb\mLst\sqrt{3q^2}} \bigg\{\frac{1}{\mLb} \left[(1+ C_0^{(v)}) (-\mLst^2+\mLst\mLb+q^2)+\frac{1}{2}C_1^{(v)} (-\mLst^2+\mLb^2+q^2)\right]\zeta_2 \nonumber \\
+&\left[(1+ C_0^{(v)})(\mLb-\mLst)+C_1^{(v)}\frac{-\mLst^2+\mLb^2+q^2}{2\mLb}\right]\zeta_1 +\left[\frac{\mLst^2-q^2}{\mLb^2}\zeta_1^\text{SL}+\zeta_2^\text{SL}\right]\nonumber\\
+&\frac{1}{2\mLb^2}\left[(2\mLst^2- \mLst\mLb-\mLb^2-2 q^2)\zeta_3^\text{SL}-(\mLst^2+ \mLst\mLb-2\mLb^2-q^2)\zeta_4^\text{SL}\right]\bigg\}\,,\label{eq:HAVHQE:3} \\
\mathcal{A}_V(+1/2, -1/2, -1) &= \frac{2\sqrt{\sPlus}}{\sqrt{3}\mLb\mLst}\bigg\{\sMinus(1+C_0^{(v)})(\zeta_2-\zeta_1)+\mLst(\zeta_1^\text{SL}+\zeta_2^\text{SL})-\frac{\sMinus}{2\mLb}(\zeta_3^\text{SL}+\zeta_4^\text{SL})\bigg\} \,, \label{eq:HAVHQE:4} 
\end{align}
and for the axial vector current:
\begin{align}
\mathcal{A}_A(+1/2, +3/2, +1)&=  2 \frac{\sMinus}{\mLb} (\zeta_1^\text{SL}-\zeta_2^\text{SL})\,, \label{eq:HAAHQE:1} \\
\mathcal{A}_A(+1/2, +1/2,  0) &= \frac{\sqrt{2\sMinus}}{\sqrt{3 q^2} \mLb\mLst}\bigg\{\frac{\sPlus}{\mLb}\left[(1+C_0^{(v)})(-\mLst^2+\mLst \mLb+q^2)+\frac{\sMinus}{2}C_1^{(v)}\right]\zeta_2 \nonumber \\
&+\sPlus\left[(\mLb-\mLst)(1+C_0^{(v)})+\frac{\sMinus}{2\mLb} C_1^{(v)}\right]\zeta_1 \nonumber\\
&+\frac{\mLb^2(\mLb^2-\mLst^2-q^2)-\sPlus\sMinus}{\mLb^2}\zeta_1^\text{SL}+(\mLst^2-\mLb^2+q^2)\zeta_2^\text{SL}\,, \nonumber \\
&-\frac{\sPlus}{2\mLb^2}\left[(2\mLst^2-3\mLst\mLb+\mLb^2-2q^2)\zeta_3^\text{SL}+(\mLst^2-3\mLst\mLb+2\mLb^2-q^2)\zeta_4^\text{SL}\right]\bigg\}\label{eq:HAAHQE:2} \\
\mathcal{A}_A(+1/2, +1/2,  t) &= \frac{\sqrt{2\sPlus}\sMinus}{\sqrt{3 q^2} \mLb\mLst}\bigg\{\frac{1}{\mLb}\left[(\mLst^2+\mLst\mLb-q^2)(1+C_0^{(v)})+\frac{1}{2}(\mLb^2-\mLst^2+q^2)C_1^{(v)}\right]\zeta_2  \,, \nonumber \\
&+\left[(\mLb+\mLst)(1+C_0^{(v)})+\frac{\mLb^2-\mLst^2+q^2}{2\mLb}C_1^{(v)}\right]\zeta_1 +\frac{\mLst^2-q^2}{\mLb^2}\zeta_1^\text{SL}-\zeta_2^\text{SL}\nonumber\\
&+\frac{1}{2\mLb^2}\left[(2\mLst^2+\mLst\mLb-\mLb^2-2 q^2)\zeta_3^\text{SL}+(\mLst^2-\mLst\mLb-2\mLb^2-q^2)\zeta_4^\text{SL}\right]\label{eq:HAAHQE:3}\\
\mathcal{A}_A(+1/2, -1/2, -1) &=\frac{2\sqrt{\sMinus}}{\sqrt{3}\mLb\mLst}\bigg\{\sPlus(1+C_0^{(v)})(\zeta_1+\zeta_2)+\mLst (\zeta_1^\text{SL}-\zeta_2^\text{SL})+\frac{\sPlus}{2\mLb}(\zeta_3^\text{SL}-\zeta_4^\text{SL})\bigg\} \,.\label{eq:HAAHQE:4}
\end{align}
In the case of the tensor current, the non-zero helicity amplitudes in the heavy quark limit are 
\begin{align}
    \mathcal{A}_T(+1/2, +3/2, +1) &=  \frac{2 \sqrt{\sPlus}}{\mLb}\bigg\{(\mLb-\mLst)\zeta_1^\text{SL}+\frac{\mLst\mLb-\mLst^2+q^2}{\mLb}\zeta_2^\text{SL}\bigg\}\,, \label{eq:HATHQE:1}\\
    \mathcal{A}_T(+1/2, +1/2,  0) & =  \frac{\sqrt{2q^2\sPlus}}{\sqrt{3}\mLb\mLst}\bigg\{\sMinus(\zeta_2-\zeta_1)+\frac{\mLst^2+\mLb^2-q^2}{\mLb}(\zeta_2^\text{SL}+\zeta_1^\text{SL})+\frac{\sMinus}{2\mLb}(\zeta_3^\text{SL}+\zeta_4^\text{SL})\bigg\}\,, \label{eq:HATHQE:2}\\
    \mathcal{A}_T(+1/2, -1/2, -1) & =\frac{2\sqrt{\sPlus}}{\sqrt{3}\mLb\mLst}\bigg\{\sMinus\left[(\mLst+\mLb)\zeta_1+\frac{\mLst^2+\mLst\mLb-q^2}{\mLb}\zeta_2\right]\nonumber\\
&+\frac{\mLst(\mLst\mLb-\mLst^2+q^2)}{\mLb}\zeta_1^\text{SL} +\mLst(\mLb-\mLst)\zeta_2^\text{SL}\nonumber \\
&+\frac{\sMinus}{2\mLb}\left[(\mLb+\mLst)\zeta_3^\text{SL}+\frac{\mLst^2+\mLst\mLb-q^2}{\mLb}\zeta_4^\text{SL}\right]\,, \label{eq:HATHQE:3}
\end{align}
and for the tensor axial current:
\begin{align}
    \mathcal{A}_{T5}(+1/2, +3/2, +1) &=-\frac{2\sqrt{\sMinus}}{\mLb}\bigg\{(\mLb+\mLst)\zeta_1^\text{SL}+\frac{\mLst^2+\mLst\mLb-q^2}{\mLb}\zeta_2^\text{SL}\bigg\} \,, \label{eq:HAT5HQE:1}\\
    \mathcal{A}_{T5}(+1/2, +1/2,  0) & = \frac{\sqrt{2 q^2 \sMinus}}{\sqrt{3}\mLb\mLst}\bigg\{ \sPlus(\zeta_1+\zeta_2)+\frac{\mLst^2+\mLb^2-q^2}{\mLb}(-\zeta_1^\text{SL}+\zeta_2^\text{SL})+\frac{\sPlus}{2\mLb}(-\zeta_3^\text{SL}+\zeta_4^\text{SL})\bigg\}\,, \label{eq:HAT5HQE:2}\\
    \mathcal{A}_{T5}(+1/2, -1/2, -1) & =  \frac{2\sqrt{  \sMinus}}{\sqrt{3}\mLb\mLst}\bigg\{\sPlus\left[(\mLb-\mLst)\zeta_1+\frac{\mLst\mLb-\mLst^2+q^2}{\mLb}\zeta_2\right]\nonumber \\
    &+\frac{\mLst(\mLst^2+\mLst\mLb-q^2)}{\mLb}\zeta_1^\text{SL}+\mLst(\mLst+\mLb)\zeta_2^\text{SL} \nonumber\\
    &+\frac{\sPlus}{2\mLb}\left[-(\mLb-\mLst)\zeta_3^\text{SL}+\frac{\mLst\mLb-\mLst^2+q^2}{\mLb}\zeta_4^\text{SL}\right]\,. \label{eq:HAT5HQE:3}
\end{align}

\section{Relations with lattice form factors}
\label{app:B}
The definitions of the form factors here employed differ from other conventions in the literature. In particular, the translation with the Lattice determination in Ref.~\cite{Meinel:2020owd} is needed. I find 
\begin{equation}
\begin{aligned}
F_{1/2,t} =&\, \frac{1}{2}\sqrt{\frac{\sMinus}{4\mLb \mLst}} f_0\,, & F_{1/2,0} =&\, \frac{1}{2}\sqrt{\frac{\sPlus}{4\mLb \mLst}} f_+\,, \\
F_{1/2,\perp} =&\, \frac{1}{2}\sqrt{\frac{\sPlus}{4\mLb \mLst}} f_\perp\,, & F_{3/2,\perp} =&\, -\frac{1}{2}\sqrt{\frac{\sPlus}{4\mLb \mLst}} f_{\perp^\prime}\,, \\
G_{1/2,t} =&\, \frac{1}{2} \sqrt{\frac{\sPlus}{4\mLb\mLst}} g_0\,, & F_{1/2,0} =&\, \frac{1}{2} \sqrt{\frac{\sMinus}{4\mLb\mLst}} g_+\,, \\
G_{1/2,\perp} =&\, \frac{1}{2} \sqrt{\frac{\sMinus}{4\mLb\mLst}} g_\perp\,, & G_{3/2,\perp} =&\, \frac{1}{2} \sqrt{\frac{\sMinus}{4\mLb\mLst}} g_{\perp^\prime}\,, \\
T_{1/2,0} =&\,  \sPlus^{1/2} \sqrt{\frac{\mLst}{4\mLb}}h_+\,,  \\
T_{1/2,\perp} =&\,  \sPlus^{1/2} \sqrt{\frac{\mLst}{4\mLb}} h_\perp\,,  & T_{3/2,\perp} =&\, \sqrt{\frac{\sPlus}{4\mLb\mLst}}(\mLb+\mLst) h_{\perp^\prime}\,,\\
T^5_{1/2,0} =&\,  \sMinus^{1/2} \sqrt{\frac{\mLst}{4\mLb}}\tilde{h}_+\,,  \\
T^5_{1/2,\perp} =&\,  \sMinus^{1/2} \sqrt{\frac{\mLst}{4\mLb}} \tilde{h}_\perp\,,  & T^5_{3/2,\perp} =&\,- \sqrt{\frac{\sMinus}{4\mLb\mLst}}(\mLb-\mLst)\tilde{h}_{\perp^\prime}\,.
\end{aligned}
\end{equation}

\section{Correlations between the fit parameters}
\label{app:C}
The correlation matrix for the HQE parameters is reported in \Table{tab:correlation}. The order of the various correlation coefficients is the same as in \Table{tab:fit_values}.
\begin{table}
\begin{center}
\renewcommand{\arraystretch}{1.2} 
 \resizebox{1\textwidth}{!}{  
\begin{tabular}{ c c c c c c c c c c }
\toprule
$1$ & $-0.879$ & $0.440$ & $0.0458$ &  $0.0460$ & $-0.120$ & $-0.0619$ & $0.363$ & $0.337$ & $0.0312$ \\
$-0.879$ & $1$ & $-0.160$ & $0.0109$ & $-0.0585$ & $0.130$ & $0.0936$ & $-0.325$ & $-0.343$ & $-0.121$ \\
$0.440$ & $-0.160$ & $1$ & $-0.00723$ & $-0.00712$ & $0.0101$ & $0.0465$ & $0.211$ & $0.139$ & $-0.218$ \\
$0.0458$ & $0.0109$ & $-0.00723$ & $1$ & $0.861$ & $-0.512$ & $-0.382$ & $-0.221$ & $-0.0500$ & $0.611$ \\
$0.0460$ & $-0.0585$ & $-0.00712$ & $0.861$ & $1$ & $-0.406$ & $-0.435$ & $-0.214$ & $-0.0603$ & $0.707$ \\
$-0.120$ & $0.130$ & $0.0101$ & $-0.512$ & $-0.406$ & $1$ & $0.887$ & $0.0941$ & $-0.119$ & $-0.783$ \\
$-0.0619$ & $0.0936$ & $0.0465$ & $-0.382$ & $-0.435$ & $0.887$ & $1$ & $0.160$ & $-0.0287$ & $-0.871$ \\
$0.363$ & $-0.325$ & $0.211$ & $-0.221$ & $-0.214$ & $0.0941$ & $0.160$ & $1$ & $0.966$ & $-0.164$ \\
$0.337$ & $-0.343$ & $0.139$ & $-0.0500$ & $-0.0603$ & $-0.119$ & $-0.0287$ & $0.966$ & $1$ & $0.0606$ \\
$0.0312$ & $-0.121$ & $-0.218$ & $0.611$ & $0.707$ & $-0.783$ & $-0.871$ & $-0.164$ & $0.0606$ & $1$ \\
\toprule
\end{tabular}
}
\caption{Correlation matrix for the HQE parameters.}
\label{tab:correlation}
\end{center}
\end{table}

\bibliographystyle{utphys}
\bibliography{references}

\end{document}